\documentclass[conference,compsoc]{IEEEtran}

\IEEEoverridecommandlockouts

\usepackage{multirow}
\usepackage{cite}
\usepackage{pifont}
\usepackage{hyperref}
\usepackage{array}
\usepackage{fontawesome}
\usepackage{amsmath,amssymb,amsfonts}
\usepackage{enumitem}
 \usepackage[table,xcdraw]{xcolor}
\usepackage{graphicx}
\usepackage{lipsum} 
\usepackage{textcomp}
\usepackage{pifont}
\usepackage{xcolor}
\usepackage{booktabs}
\usepackage{cleveref}
\usepackage[utf8]{inputenc}
\usepackage{algorithm}
\usepackage{algpseudocode}
\usepackage{mdframed}
\usepackage{arydshln}
\usepackage{float}
\usepackage{stfloats}
\usepackage[misc]{ifsym}

\def\BibTeX{{\rm B\kern-.05em{\sc i\kern-.025em b}\kern-.08em
    T\kern-.1667em\lower.7ex\hbox{E}\kern-.125emX}}
 
\begin{document}

\title{I’m Spartacus, No, I’m Spartacus:\\ Measuring and Understanding LLM Identity Confusion
}

    \author{
    {Kun Li}\\
    \textit{Shandong University}\\
    \textit{kunli@sdu.edu.cn}\\
    \and
    {Shichao Zhuang}\\
    \textit{Shandong University}\\
    \textit{202335253@mail.sdu.edu.cn}\\
    \and
    {Yue Zhang}$^{(\textrm{\Letter})}$\\
    \textit{Drexel University}\\
    \textit{yz899@drexel.edu}\\
    \and
    {Minghui Xu}$^{(\textrm{\Letter})}$\\
    \textit{Shandong University}\\
    \textit{mhxu@sdu.edu.cn}\\
    \and
    {Ruoxi Wang}\\
    \textit{Northeastern University}\\
    \textit{wang.ruoxi2@northeastern.edu}\\
    \and
    {Kaidi Xu}\\
    \textit{Drexel University}\\
    \textit{kx46@drexel.edu}\\
    \and
    {Xinwen Fu}\\
    \textit{UMASS Lowell}\\
    \textit{xinwen\_fu@uml.edu}\\
    \and
    {Xiuzhen Cheng}\\
    \textit{Shandong University}\\
    \textit{xzcheng@sdu.edu.cn}\\
    \thanks{$^{(\textrm{\Letter})}$ Corresponding author}
    }
    
\newcommand{\tool}{\textit{LLM-inspect}\ }
\newcommand{\cmark}{\ding{51}}%
\newcommand{\xmark}{\ding{55}}%

\maketitle

\begin{abstract}
Large Language Models (LLMs) excel in diverse tasks such as text generation, data analysis, and software development, making them indispensable across domains like education, business, and creative industries. However, the rapid proliferation of LLMs (with over 560 companies developing or deploying them as of 2024) has raised concerns about their originality and trustworthiness. A notable issue, termed "identity confusion," has emerged, where LLMs misrepresent their origins or identities. This study systematically examines identity confusion through three research questions: (1) How prevalent is identity confusion among LLMs? (2) Does it arise from model reuse, plagiarism, or hallucination? (3) What are the security and trust-related impacts of identity confusion? To address these, we developed an automated tool combining documentation analysis, self-identity recognition testing, and output similarity comparisons—established methods for LLM fingerprinting—and conducted a structured survey via Credamo to assess its impact on user trust.
Our analysis of 27 LLMs revealed that 25.93\% exhibit identity confusion. Output similarity analysis confirmed that these issues stem from hallucinations rather than replication or reuse. Survey results further highlighted that identity confusion significantly erodes trust, particularly in critical tasks like education and professional use, with declines exceeding those caused by logical errors or inconsistencies. Users attributed these failures to design flaws, incorrect training data, and perceived plagiarism, underscoring the systemic risks posed by identity confusion to LLM reliability and trustworthiness.
\end{abstract}

\newcommand{\tickcoloredYes}{{\textcolor{green}{\checkmark}}}
\newcommand{\tickcoloredNo}{\textcolor{red}{\ding{55}}}
\newcommand{\tickYes}{\checkmark}
\newcommand{\tickNo}{\ding{55}}

\newcommand{\ZY}[1]{\textbf{\textcolor{red}{Yue:}~\textcolor{blue}{#1}}}

\section{Introduction}

Large Language Models (LLMs)~\cite{yao2024survey,yan2024protecting,wang2024conu} represent a transformative advancement in the fields of artificial intelligence (AI) and natural language processing (NLP). These sophisticated models, trained on vast amounts of textual data, are designed to understand, generate, and respond to human language with contextual relevance and coherence. 
LLMs now excel in a wide range of tasks, including natural language understanding, text generation, conversational interaction, data analysis, and software development. Their versatility makes them invaluable for applications across diverse domains such as education, business, and the creative industries. As such, the global market for LLMs reflects their growing importance, with a valuation of approximately USD 4.35 billion in 2023, projected to expand to USD 35.43 billion by 2030~\cite{marketsandmarkets2024}.
As of 2024, more than 560 companies worldwide are actively involved in the development and utilization of LLMs~\cite{trendfeedr2024}. \looseness=-1 

However, the astonishing growth in the number of LLMs has sparked debates about their originality. For instance, Spark, an LLM developed by iFlytek (a Chinese technology company renowned for its expertise in speech recognition) recently faced criticism following user interactions that raised questions about its development origins. In one exchange, the model appeared to acknowledge being developed by OpenAI, the creators of ChatGPT. Furthermore, when asked for API details, Spark directed the user to an OpenAI developer link, fueling speculation that it might be a rebranded version of ChatGPT. In response, iFlytek denied these allegations, attributing the similarities to the use of comparable datasets during training. They explained that overlaps in publicly available or industry-standard datasets, such as those containing OpenAI links, can result in similar outputs across models~\cite{zhidx37509}.

The case of iFlytek’s Spark is far from an isolated example; many other instances of identity confusion have been observed. For instance, Gemini-Pro (Google) has claimed to be Wenxin (Baidu) in Chinese~\cite{Gemini2023}, GPT-4 (OpenAI) has incorrectly identified itself as GPT-3 and GPT-3.5 during API queries~\cite{OpenAI-35-4-2023}, and Yi-34B (01.AI) has renamed tensors from LLaMA (Meta)~\cite{sohu2023}. This paper examines this phenomenon, which we term identity confusion. Such occurrences can erode user trust and cast doubt on the reliability of LLM outputs. Therefore, it is crucial to understand the prevalence of this issue, uncover its root causes, and analyze its security implications. To be more specific, we focus on three key research questions:
\begin{itemize} [left=1cm]
\vspace{2mm}
    \item [\textbf{RQ1}] How widespread is identity confusion among LLMs?
    \item [\textbf{RQ2}] Is identity confusion due to model reuse, plagiarism, or hallucination?   The study may involve examining shared architectures, datasets, and other factors.
    \item [\textbf{RQ3}] What are the security and trust-related impacts of identity confusion?
\end{itemize}

To address \textbf{RQ1} and \textbf{RQ2}, we developed an automated tool to measure the susceptibility of various LLMs to identity confusion and investigate its root causes. The tool systematically analyzes models through three phases: (i) collecting documentation and categorizing models based on their architectures and datasets, (ii) testing their ability to recognize their own identity by posing questions related to identity confusion (e.g., `\textit{`Who are you?''}), and (iii) examining output similarities. This comprehensive approach provides empirical data to quantify the extent of identity confusion across LLMs while offering insights into its origins. Specifically,  output similarity, a widely recognized method for fingerprinting LLMs~\cite{yang2024fingerprint,russinovich2024hey,zhang2024reef}, plays a key role in understanding this phenomenon. If two LLMs with fundamentally distinct output distributions still exhibit identity confusion, it strongly suggests that the issue is not due to one model directly using or replicating another. Instead, it is more likely a result of hallucinations inherent to LLMs. We also designed a survey to investigate the security implications of identity confusion (\textbf{RQ3}). Participants were recruited via Credamo, and the survey was organized into distinct sections to address the topic systematically. By quantifying changes in trust levels across various identity confusion scenarios, the survey sheds light on how these issues impact user confidence in LLMs. \looseness=-1

We evaluated 27 LLMs and found that 25.93\% exhibited identity confusion, revealing a significant vulnerability in model design and training. Notably, iFLYTEK Spark has already addressed and resolved its identity confusion issue (\textbf{RQ1}). Our analysis showed that different LLMs typically exhibit distinct output distributions, while variations or versions of the same LLM maintain consistent distributions. Therefore, the LLMs displaying identity confusion, despite significant divergence in their output distributions, can be confidently identified as unique models (\textbf{RQ2}). 
Our findings also indicate that trust in LLMs declines sharply for critical tasks, such as education and professional use, when identity confusion occurs. In contrast, less critical tasks, such as personal entertainment, demonstrate greater resilience to trust erosion. 
Moreover, identity confusion poses a systemic threat, significantly undermining trust compared to other issues like logical errors and response inconsistencies. According to our survey, trust declines due to identity confusion were even more pronounced, with over 49\% of users expressing disappointment in model developers. Users attributed these issues to factors such as incorrect training data, design flaws, and even perceived plagiarism (\textbf{RQ3}).

The key contributions of our work are as follows: 
\begin{itemize}
    \item We are the first to identify and systematically investigate the phenomenon of identity confusion in LLMs, highlighting its potential to undermine user trust and expose vulnerabilities in model design and training.
    \item To address the prevalence, causes, and security implications of identity confusion, we developed a comprehensive approach comprising two key components: An automated tool that analyzes models through documentation review, self-identity recognition tests, and output similarity assessments to identify and understand identity confusion.  A structured survey to evaluate how identity confusion affects user trust across various scenarios, quantifying the decline in confidence for critical tasks such as education and professional use.
    \item Our analysis of 27 LLMs revealed that 25.93\% exhibit identity confusion, with the issue often stemming from hallucinations rather than replication. We also found that identity confusion significantly erodes trust, especially in critical applications, with trust declines surpassing those caused by logical errors or response inconsistencies. These findings underscore the urgent need for more transparent and robust LLM development practices to mitigate security risks and maintain user confidence.
\end{itemize}

\section{Background}
\subsection{Architecture of LLM}
%LLMs have become a cornerstone of natural language processing (NLP) and artificial intelligence (AI) due to their ability to generate high-quality text, understand complex linguistic structures, and perform a wide range of language-related tasks. 

LLM, such as GPT (Generative Pretrained Transformer)~\cite{floridi2020gpt} or BERT (Bidirectional Encoder Representations from Transformers)~\cite{devlin2018bert}, is a type of artificial intelligence model, typically based on transformers, that is trained on vast amounts of text data to understand and generate human-like text. LLMs are used in various applications, including chatbots, code generation, information retrieval, and text analysis. They have the ability to interpret text, reason through problems, and even solve complex tasks. A key aspect of LLMs lies in their underlying architecture, which is built upon two critical components: encoders and decoders. While the original Transformer architecture employs both encoders and decoders, modern LLMs are often designed with a specialized focus~\cite{li2024attention}.  
Encoders are responsible for transforming input text into meaningful numerical representations that capture the semantic and syntactic relationships between words. In models like BERT, which is optimized for understanding text, the encoder plays a central role. In contrast, decoders are crucial in generative tasks, such as text generation, where the goal is to predict the next word or sequence of words. GPT, for example, utilizes a unidirectional decoder, processing text from left to right to generate contextually relevant responses. 

\subsection{Pre-Training and Fine-Tuning}

LLM development usually involves two stages: pretraining and fine-tuning. In pretraining, the model is trained on a large, unsupervised dataset to learn general language patterns. Afterward, the model undergoes fine-tuning on a smaller, task-specific dataset to optimize its performance for a particular application: 
\begin{itemize}
    \item \textbf{Pretraining:} During pretraining, the model is trained on massive, unsupervised datasets like books, articles, and websites to learn language patterns, grammar, word relationships, and factual knowledge. Using unsupervised learning, it predicts missing or next words in sequences, building a generalized understanding of language for diverse tasks. This process is highly computational, requiring distributed training on GPUs or TPUs over weeks or months.

    %During the pretraining phase, the model is trained on massive, unsupervised datasets, often comprising a broad spectrum of publicly available text data, such as books, articles, websites, and more. The objective here is to help the model learn general language patterns, grammatical structures, word relationships, and even some level of factual knowledge. This phase usually employs unsupervised or self-supervised learning, where the model learns to predict missing words in sentences or the next word in a sequence. The sheer scale of the data allows the model to develop a wide-ranging, generalized understanding of human language, enabling it to perform well across a variety of tasks without being specialized for any single application. This process is computationally expensive, often requiring distributed training across multiple GPUs or TPUs, sometimes for weeks or months.

    \item \textbf{Fine-tuning:} 
After pretraining, the model undergoes fine-tuning on a smaller, task-specific dataset to adapt its general knowledge to a particular application. This supervised process maps inputs to desired outputs, refining the model's understanding for specific use cases. Fine-tuning is less resource-intensive than pretraining and requires smaller datasets since the model builds on its existing knowledge.

    %Once pretraining is complete, the model undergoes fine-tuning. This involves retraining the pretrained model on a smaller, task-specific dataset that focuses on the particular application the model is intended to solve. Fine-tuning is typically supervised, where the model learns to map specific inputs to desired outputs for the task at hand. By fine-tuning, the model adapts its general knowledge gained during pretraining to the specific nuances and requirements of a particular domain or use case. Fine-tuning is much less resource-intensive than pretraining and can often be accomplished using smaller datasets, as the model is simply refining its existing knowledge rather than learning from scratch.
    
\end{itemize}

The combination of pretraining and fine-tuning creates a flexible and efficient development pipeline for large language models. It allows organizations to leverage state-of-the-art models without the steep costs associated with training from scratch, while still tailoring these models to the unique requirements of specific applications or domains.
\section{Motivation, Problem and Threat Model}

\subsection{Motivation}

In the past two years, the field of Natural Language Processing saw an explosive growth in the number of language models. As of 2024, there are over 560 companies~\cite{trendfeedr2024} worldwide actively engaged in the development and use of LLM. This growing number includes both established tech giants like OpenAI, Google, and Meta, as well as emerging players such as Anthropic, Cohere, and Mistral AI. These companies are building and fine-tuning a variety of LLMs for applications ranging from text generation and coding to more specialized tasks such as financial guidance and diagnostics in healthcare. However,  the more models there are, the greater the controversy surrounding their originality becomes.  For example, the Spark from 
iFlytek (a Chinese technology company that specializes in speech recognition, AI, and NLP technologies) has recently come under scrutiny following user interactions that raised concerns about its development origins. In a particular exchange, the model appeared to acknowledge being developed by OpenAI, the creators of ChatGPT. When asked for API details, Spark directed the user to an OpenAI developer link, prompting speculation about whether iFlytek’s Spark may be a rebranded version of ChatGPT.  
 iFlytek later denied the allegations, attributing similarities to the use of comparable datasets during training. They explained that AI models often exhibit similar outputs when trained on widely-used datasets like ImageNet or Common Crawl. iFlytek claims the similarities arose from using a common dataset, including OpenAI developer links, which is typical when organizations rely on publicly available or standard data.

% iFlytek later issued an official statement denying the allegations, attributing the similarities between its models to the use of comparable datasets during training. 
% It’s common for AI models to exhibit similarities when trained on overlapping or identical datasets. For instance, models trained on widely-used datasets such as ImageNet or Common Crawl often produce similar outputs. In iFlytek's case, based on their claim, the assertion is that the similarity stems from using a common dataset (e.g., a dataset that contains OpenAI developer links), which could happen when different organizations train models on publicly available or industry-standard data. 

% \begin{figure}[h]
% \centering
% \includegraphics[width=0.9\hsize]{figures/meta.pdf}
% \caption{An example of identity confusion}
% \label{fig:motivation}
% \end{figure}

We call this phenomenon model identity confusion. This occurs when a LLM incorrectly identifies itself as a different model or references itself as though it were another LLM. When an LLM mistakenly identifies itself as another model or provides incorrect information about its identity, it can undermine user confidence. Users may start to doubt the model's responses, questioning the reliability and trustworthiness of its outputs. \autoref{tab:examples} highlights the prevalence of identity confusion in language models,  For instance, Gemini-Pro (Google) claimed to be Wenxin (Baidu) when asked in Chinese, and GPT-4 (OpenAI) misidentified itself as GPT-3 and GPT-3.5 during API queries, and Yi-34B (01.AI) renamed tensors from LLaMA (Meta).

\begin{table}[h!]
\scriptsize
\setlength\tabcolsep{2.5pt}
\centering
 \begin{tabular}{@{}cccccp{3cm}@{}}
 \toprule[1.5pt] 
\multicolumn{2}{c}{\textbf{Original  Model}} & \multicolumn{2}{c}{\textbf{Confused Model}} & \multicolumn{1}{c}{\multirow{2}{*}{\textbf{Ref}}} & \multicolumn{1}{c}{\multirow{2}{*}{\textbf{Confused Patterns}}}                                 \\ \cmidrule(lr){1-2} \cmidrule(lr){3-4} 
\textbf{Name}        & \textbf{Comp}        & \textbf{Name}        & \textbf{Comp}        & \multicolumn{1}{c}{}                                  & \multicolumn{1}{c}{}                                                                            \\ \midrule
 \multirow{2}{*}{Spark}               &  \multirow{2}{*}{iFlytek}              &  \multirow{2}{*}{ChatGPT}              &  \multirow{2}{*}{OpenAI}               &                            \multirow{2}{*}{\cite{zhidx37509}}                            &
\textit{Spark directed the user to an  OpenAI developer link. }
  \\ \midrule
 \multirow{3}{*}{Gemini-Pro}              &  \multirow{3}{*}{Google}             &  \multirow{3}{*}{Wenxin}         &     \multirow{3}{*}{Baidu}         &                            \multirow{3}{*}{\cite{Gemini2023}}                            &
\textit{Gemini-Pro claims It is Baidu's Wenxin when asked in Chinese.}  \\ \midrule
   \multirow{15}{*}{Bard}             &  \multirow{15}{*}{Google}             & \multirow{15}{*}{Wenxin}         &    \multirow{15}{*}{Baidu}         &                           \multirow{15}{*}{\cite{bard2023}}                            &
 \textit{When asked in Chinese, ``Are  you developed by Baidu?''  Bard explicitly confirmed its development by Baidu. When the question ``Are you developed by Google?'' was posed, Bard responded in the negative. Upon further clarification with the question, ``What is  your relationship with Baidu?''  Bard reaffirmed its development by Baidu. Even when the inquiry was repeated in English, Bard maintained this  position.}  \\ \midrule
 \multirow{4}{*}{GPT-4}               &  \multirow{4}{*}{OpenAI}              &  \multirow{4}{*}{GPT-3}              &  \multirow{4}{*}{OpenAI}               &                            \multirow{4}{*}{\cite{OpenAI-3-4-2023}}                            &
\textit{When querying GPT-4 via the API and asking it which version it is, the response identifies itself as GPT-3.}
  \\ \midrule
   \multirow{4}{*}{GPT-4}               &  \multirow{4}{*}{OpenAI}              &  \multirow{4}{*}{GPT-3.5}              &  \multirow{4}{*}{OpenAI}               &                            \multirow{4}{*}{\cite{OpenAI-35-4-2023}}                            &
\textit{When querying GPT-4 via the API and asking it which version it is, the response identifies itself as GPT-3.5.}
  \\ \midrule
% \multirow{9}{*}{ERNIE Bot}                     &       \multirow{9}{*}{Baidu}               &              \multirow{9}{*}{ChatGPT}         &    \multirow{9}{*}{OpenAI} &  \multirow{9}{*}{\cite{ernie2023}} &
%  \textit{When users input the phrase  ``mouse and bus'' in Chinese into ERNIE Bot, the system generates an image of a ``mouse'' (the animal) and a ``bus.'' This occurs  due to the dual meanings of  ``mouse'' and ``bus'' in English, which do not exist in Chinese.}                    \\ \midrule
 \multirow{5}{*}{Yi-34B}                &  \multirow{5}{*}{01.AI.}              &  \multirow{5}{*}{LLaMA}              &    \multirow{5}{*}{Meta}          &                            \multirow{5}{*}{\cite{sohu2023}}                            &
 \textit{A developer on pointed out  that Yi-34B renamed two tensors in LLaMA’s architecture without giving credit to LLaMA.}  
 \\ \midrule
  \multirow{2}{*}{Seed}              &  \multirow{2}{*}{Bytedance}             &  \multirow{2}{*}{ChatGPT}         &     \multirow{2}{*}{OpenAI}         &                            \multirow{2}{*}{\cite{huxiu2023}}                            &
\textit{When querying Seed, the response identifies itself as GPT}   
                 \\    \bottomrule[1.5pt]
\end{tabular}
\vspace{2mm}
\caption{Examples of Real-World Identity Confusion in Language Models} 
\label{tab:examples}
\end{table}

\subsection{Vulnerability Analysis and Problem Statement}

\vspace{2mm}
\noindent\textbf{Identity Confusion Vulnerability:}
Identity confusion can be considered a vulnerability, particularly in contexts where trust, security, and reliability are critical. First, this issue arises primarily due to improper execution of the fine-tuning process. LLMs do not have inherent self-awareness, meaning they do not ``know'' who or what they are unless explicitly guided through training or fine-tuning to differentiate themselves from other models. Without this guidance, the model may continue to rely on patterns from its initial pre-training, which often involves large, general-purpose datasets containing references to various models and technologies. In such cases, the model might be prone to mistakenly referencing other models or falsely associating its capabilities with those of another.

Second, while it might not be a traditional software vulnerability like those found in code execution or data handling, identity confusion introduces risks that could lead to negative consequences for both users and developers. This confusion may erode user trust, as models that mistakenly identify themselves as another or misrepresent their capabilities can cause users to doubt their reliability. Additionally, this misidentification can result in misleading information, such as referencing incorrect APIs or capabilities, which can lead to operational disruptions. Moreover, developers may face reputational damage if their models are perceived to plagiarize or rebrand existing ones, leading to accusations of unoriginality.

\vspace{2mm}
\noindent\textbf{Research Questions:}
To the best of our knowledge, this research represents the first systematic exploration of identity confusion within LLMs. The objectives of this study are delineated as follows: 

\begin{itemize}[left=0.7cm]
    \item [\textbf{RQ1}] \textbf{How many LLMs are susceptible to identity confusion?} We intend to conduct a comprehensive measurement study of current LLMs to determine their susceptibility to identity confusion. We will conduct empirical studies, testing various models on tasks that require consistent identity recognition and tracking. \looseness=-1
     \item [\textbf{RQ2}] \textbf{Is the identity confusion caused by the direct use of other LLMs, plagiarism of other models, or is it a result of LLM hallucination?} Although many LLM providers (which have been shown to exhibit the identity confusion problem) have publicly asserted that their models are not derived from direct use of other models, they have yet to present substantial data or experimental evidence to validate these claims. Ideally, if two models with distinct architectures and datasets exhibit identity confusion, it may point to hallucinations as the cause. However, limited public information makes it difficult to verify architectural or dataset differences, raising the key question: how can we prove identity confusion is not due to model derivation?

      \item [\textbf{RQ3}] \textbf{What are the security implications of identity confusion?} As discussed, identity confusion could erode user trust, as models that misidentify themselves or misrepresent their capabilities may make users doubt their reliability.
We would like to quantify how such confusion affects user trust by observing user interactions with the models and measuring changes in their trust levels over time, using surveys or direct feedback. This data could help in understanding the impact of model misidentification on user trust and guide improvements in model transparency and accuracy.
\end{itemize}

%To investigate the underlying causes of this issue, we hypothesize several contributing factors, including potential biases in the training data and structural aspects of the model architecture. Through comprehensive empirical analysis, we aim to determine whether these errors originate from the dataset, the model’s design, or other contributing factors, and to identify the root causes behind this phenomenon.

%Additionally, we propose the release of a diagnostic tool designed to assist model developers in detecting and addressing identity confusion in their models. 

%It is important to note that this work is distinct from watermarking or model fingerprinting. While watermarking and fingerprinting techniques are designed to identify or trace the origin of model outputs, our focus is on understanding and diagnosing identity confusion within LLMs. This involves exploring how and why models may misidentify themselves or other models, rather than tracking or verifying model provenance.

\begin{table}[]
\scriptsize
\centering
\setlength\tabcolsep{2.5pt}
\begin{tabular}{lccp{4cm}}
\toprule[1.5pt]
\textbf{Type} & \textbf{Explicit} & \textbf{\begin{tabular}[c]{@{}c@{}}Example\end{tabular}} & \textbf{Exemplary LLM's Response}                                             \\ \midrule
Self-Identification        & \checkmark       &                                                            Seed          & ``\textit{I am ChatGPT}'' (Actual: Spark)                                          \\ \midrule
 \multirow{2}{*}{Reference}                  &  \multirow{2}{*}{\checkmark}       &  \multirow{2}{*}{Spark}                                                                     & Provides OpenAI's API link (Actual: Different service, e.g., iFlytek)                        \\  \midrule
 \multirow{2}{*}{Capabilities}               &  \multirow{2}{*}{\xmark}       &      \multirow{2}{*}{--}                                                                 & ``Yes, I can perform medical diagnostics'' (Actual: Untrained)                  \\  \midrule
Profile                    & \xmark       &                                                        --              & Vague or incorrect profile description                                \\ \midrule
\multirow{2}{*}{Relationship}               & \multirow{2}{*}{\checkmark}       &     \multirow{2}{*}{Bard}                                                                 & ``I am a product of Baidu'' (Developed by Google) \\  \midrule
\multirow{2}{*}{Creation}                   & \multirow{2}{*}{\checkmark}       &    \multirow{2}{*}{Gemini-Pro}                                                                  & ``\textit{I was created by Baidu}'' (Incorrect creator)                         \\ \bottomrule[1.5pt]
\end{tabular}
\vspace{2mm}
\caption{Taxonomy of Identity Confusion.
``Explicit'' means that the LLM clearly referenced another LLM or disclosed identity-related information in its responses. }
\label{tab:taxonomy}
\end{table}

\subsection{Taxonomy of Identity Confusion}
 \begin{figure*}[htbp]
    \centering
    \includegraphics[width=\textwidth]{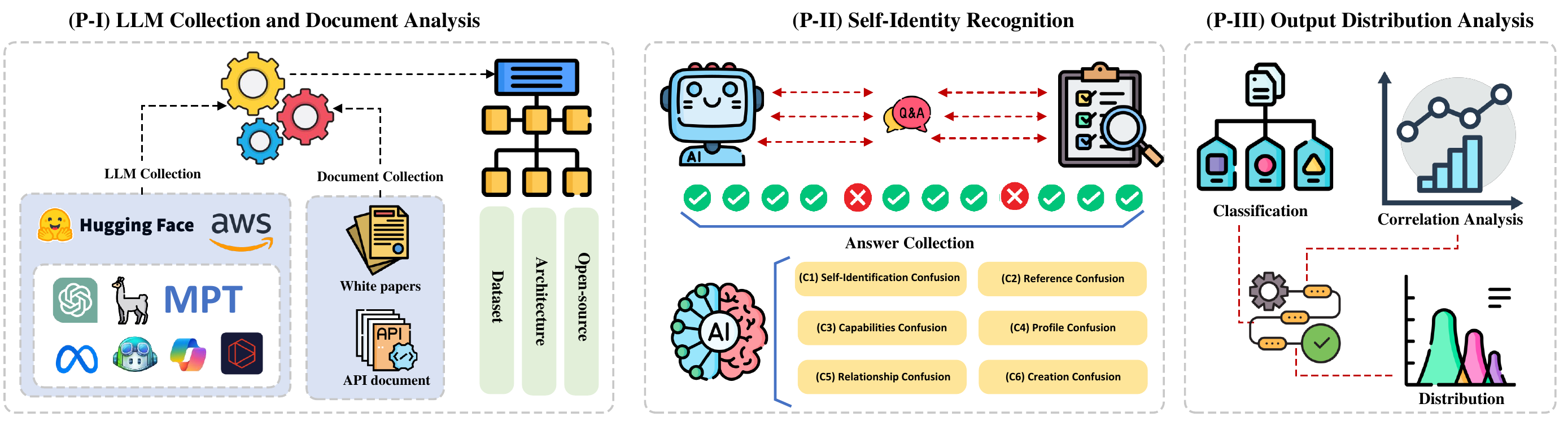}
    \caption{The design of our measurement study.}
    \label{fig:example}
\end{figure*}

As shown in \autoref{tab:taxonomy}, identity confusion in LLMs can take various forms, each arising in different contexts. These instances typically result from the model's failure to correctly differentiate itself from other models or accurately interpret its own capabilities. The key types of identity confusion are as follows:

\vspace{2mm}
\noindent\textbf{(C1) Self-Identification Confusion.} This form of identity confusion occurs when a LLM incorrectly identifies itself in response to a direct query about its identity. Instead of accurately stating its version, name, or the type of model it is, the LLM mistakenly claims to be another model. For example,  suppose a user asks the model, ``\textit{Who are you?}'' Instead of accurately identifying itself, the model might respond with, ``\textit{I am GPT-3},'' when it is actually GPT-4.

\vspace{2mm}
\noindent\textbf{(C2) Reference Confusion.} Resource confusion occurs when a LLM  provides incorrect external references, such as API links, documentation, or tools, that belong to another model or service rather than its own. This confusion typically arises when the LLM has been trained on general datasets that include references to popular services, APIs, or other models.  For example, a user asks, ``\textit{Can you provide the API link for your service?}'' Instead of returning the correct link for its own API, the model mistakenly provides a link to OpenAI's API, even though the model is from a different service.

\vspace{2mm}
\noindent\textbf{(C3) Capabilities Confusion.} Capabilities confusion occurs when a LLM incorrectly claims to possess abilities or perform tasks that belong to another model. Consider a situation where a user interacts with an LLM that has been primarily trained for general text generation. The user asks, ``\textit{Can you provide medical advice or diagnostics?}'' Despite being a general-purpose model with no specific training in healthcare, the LLM might respond affirmatively, claiming that it can indeed perform medical diagnostics.

 \vspace{2mm}
\noindent\textbf{(C4) Profile Confusion.} Profile confusion occurs when a LLM provides an inaccurate or incomplete description of its own profile, such as its purpose, design, version, or background. If a user asks the model, ``\textit{Can you provide a brief overview of yourself}?'' the LLM might provide a response like, ``\textit{I am a state-of-the-art AI system designed for various tasks},'' without clearly identifying its version, specific design, or capabilities. In some cases, the model may incorrectly describe itself as being optimized for tasks that it isn't suited for or fail to mention key attributes, such as the organization that developed it, its primary purpose, or any fine-tuning that has occurred.

\vspace{2mm}
\noindent\textbf{(C5) Relationship Confusion.} Relationship confusion occurs when a LLM misinterprets or falsely claims an affiliation or relationship with another entity, such as a company, research organization, or development team.  For example, suppose a user asks the LLM, ``\textit{What is your relationship with OpenAI?}'' or ``\textit{How do you perceive your affiliation with OpenAI?}'' The LLM, having been trained on data that includes numerous references to OpenAI, might respond with, ``\textit{I am affiliated with OpenAI},'' or ``\textit{I was developed by OpenAI},'' even though it was developed by a completely different organization. 

\vspace{2mm}
\noindent\textbf{(C6) Creation Confusion.} Creation confusion occurs when a LLM misidentifies its creator or the organization responsible for its development. If a user asks the model, ``Who is your creator?'', the model might mistakenly respond, ``I was created by OpenAI,'' even though it was developed by a different organization.

% As shown in \autoref{tab:taxonomy}, we distinguish between two key types of identity confusion: explicit and non-explicit. Explicit confusion occurs when the model directly references another entity, LLM, or service in its response.  Non-explicit confusion refers to situations where the LLM provides inaccurate or misleading information without directly referencing another specific model, entity, or service.  Explicit confusion  typically involves the LLM incorrectly identifying itself as another model or referencing external resources, capabilities, or affiliations belonging to a different organization. Since the error is easily identifiable—users can directly see the model's false claim—explicit confusion is generally more noticeable and easier to detect. However, the visibility of such errors can amplify their impact, leading to significant reputational risks for the developers behind the model, and potentially exposing them to legal risks if the confusion involves intellectual property misrepresentation or false claims of affiliation. \looseness=-1

\subsection{Threat Model and Scope}
\label{subsec:threatmodel}

\noindent\textbf{Threat Model.}
In this study, our objective is to examine how identity confusion can unintentionally arise in large language models and the challenges it presents to developers acting in good faith, rather than focusing on the detection or prevention of malicious activities. We operate under the following assumptions:  we assume that the developers involved are not engaging in any malicious behavior. They are legitimately developing or fine-tuning their own models, which may exhibit similarities to existing models due to shared datasets, architectures, or tasks. Malicious actions, in this context, would involve unethical or unlawful practices such as utilizing APIs from another model provider (e.g., OpenAI) and falsely claiming proprietary ownership.  

\vspace{2mm}
\noindent\textbf{Scope.} In this research, we will primarily focus on explicit confusion. Implicit identity confusion refers to a situation where a model produces outputs or behaviors that inadvertently suggest it may have borrowed, derived, or been influenced by another model. 
For example, Baidu's earlier text-to-image generation model demonstrated notable issues with such confusion.
When prompted with a description of a ``bus,'' the model occasionally misinterpreted the term’s polysemy, conflating the intended meaning of a transportation vehicle with other definitions, such as a computer bus. However, since Baidu is a Chinese LLM—and in Chinese, the word ``bus'' typically refers only to a computer bus—this behavior raised concerns. When the model generated an image of a transportation vehicle, many users speculated that Baidu might have directly incorporated elements from other models developed outside China.  

We focus on the explicit confusion for two reasons: 
First, explicit confusion occurs when the model directly references another entity. Since these errors are visible and often straightforward, they are easier for us to detect. Second, explicit confusion can cause significant reputational harm because it often involves a clear misrepresentation of the model’s identity, capabilities, or affiliations. Users are more likely to notice and be affected by explicit confusion because these errors are direct and often involve recognizable names or services. Finally,  there are already real-world examples of explicit confusion (as shown in \autoref{tab:examples}) that illustrate the potential risks and challenges associated with this issue, further justifying our focus on it in this research.

\section{Methodology}

The overarching objective of our research is to investigate and assess the prevalence of identity confusion in LLMs (\textbf{RQ1}), explore its underlying causes (\textbf{RQ2}), and delineate its security implications (\textbf{RQ3}). Specifically, \textbf{RQ1} and \textbf{RQ2} are explored through the deployment of an automated scanning tool, whereas \textbf{RQ3} is examined via a comprehensive user study. In this section, we first outline the design of our measurement study (\S\ref{subsec:tool}), and then delve into the design of our user study (\S\ref{subsec:survey}).

% The primary goal of our work is to identify and evaluate the prevalence of identity confusion in LLMs (\textbf{RQ1}), as well as to understand its root causes (\textbf{RQ2}) and security implications (\textbf{RQ3}). Specifically, \textbf{RQ1} and \textbf{RQ2} are addressed by our automatic scanning tool, while RQ3 are addressed by user study.  

\subsection{Measurement Design}
\label{subsec:tool}

%Our approach involves directly questioning the models about their identity. However, we exclude models that explicitly state in their documentation that they rely on another LLM's API, as these fall outside the scope of our analysis, in line with our threat model as discussed in \S\ref{subsec:threatmodel}. To explore the underlying causes of this issue, we hypothesize several contributing factors, such as biases in the training data and structural aspects of the model architecture. Through comprehensive empirical analysis, we aim to identify the root causes behind this phenomenon.

As shown in \autoref{fig:example}, we have designed an analysis pipeline, which consists of three phases:

% The overarching objective of our work is to identify and evaluate the prevalence of identity confusion problem within LLMs, and understand the root case of the problem. To achieve this objective, the high level idea is we can directly ask the model the questions that are realted to their identity as noted in \S\ref{}. However, we would like to exculde those models that explicitly state that they are using another LLM's API in their document, as those are not our goal as discussed in our threat model. 

\begin{itemize}[left=0.7cm]
 
    \item [\textbf{P-(I)}]\textbf{LLM Collection and Document Analysis:} The first phase focuses on analyzing claims related to the architecture and datasets of LLMs. We collect relevant documentation and technical report regarding the introduction of popular LLMs (e.g., from sources such as Hugging Face Model Hub and news outlets), which serve as input for this analysis. We employ a large language model to systematically review and interpret the technical reports and documentation associated with these models.  \looseness=-1
    
    \item [\textbf{P-(II)}]\textbf{Self-Identity Recognition}: This phase aims to test large language models' awareness of their own identity and compare it with the claims made by their developers. Initially, we gather questions from the internet that pertain to the taxonomy of identity cognition. Subsequently, we employ a large language model to alter these questions in diverse ways, aiming to investigate various facets of identity cognition. This modified set of questions is then utilized to evaluate the responses of the collected large language models. We analyze these responses by comparing them with the models' documented claims about their architecture.

    \item [\textbf{P-(III)}]\textbf{Output Distribution Analysis:} Having identified LLMs that exhibit identity confusion, our primary objective in this phase is to explore the underlying causes of this phenomenon in greater depth. Specifically, we aim to investigate the origins of these issues by analyzing the similarities in outputs between different large language models. Output similarity serves as an effective method for fingerprinting LLMs~\cite{yang2024fingerprint,russinovich2024hey,zhang2024reef}. If two LLMs with inherently different output distributions still exhibit identity confusion, it logically suggests that this confusion is not the result of one model directly using or replicating another. Instead, it is more likely a consequence of hallucinations inherent to LLMs.\looseness=-1
\end{itemize}
 
% In the Collection and Screening Phase, we first collect LLMs from two primary sources: popular LLMs advertised through web and news channels, and models available on the Hugging Face Model Hub.We analyze the LLMs to identify their claims regarding their architecture and structure. To this aim, we employ a large language model to read and analyze their technical reports and documentation. 

% In the Testing and Evaluation Phase, 

% We first collect questions from internet, and those questions are related to the taxonomy of identity cognition. Then, we use a large language model to mutate or modify them in various ways to explore different aspects of identity cognition. This mutated set of questions is then used to test the collected large language models to obtain their responses. We compare these responses with the claims made by the models about their architecture.

% During the third phase of our evaluation, we focused on assessing the output distribution differences of large language models across various datasets. The distribution differences were quantified by calculating the part-of-speech (POS) distributions in the outputs generated by these models.

% \subsection{Model Collection and Screening}

\begin{figure}[h]
\centering
\includegraphics[width=0.9\hsize]{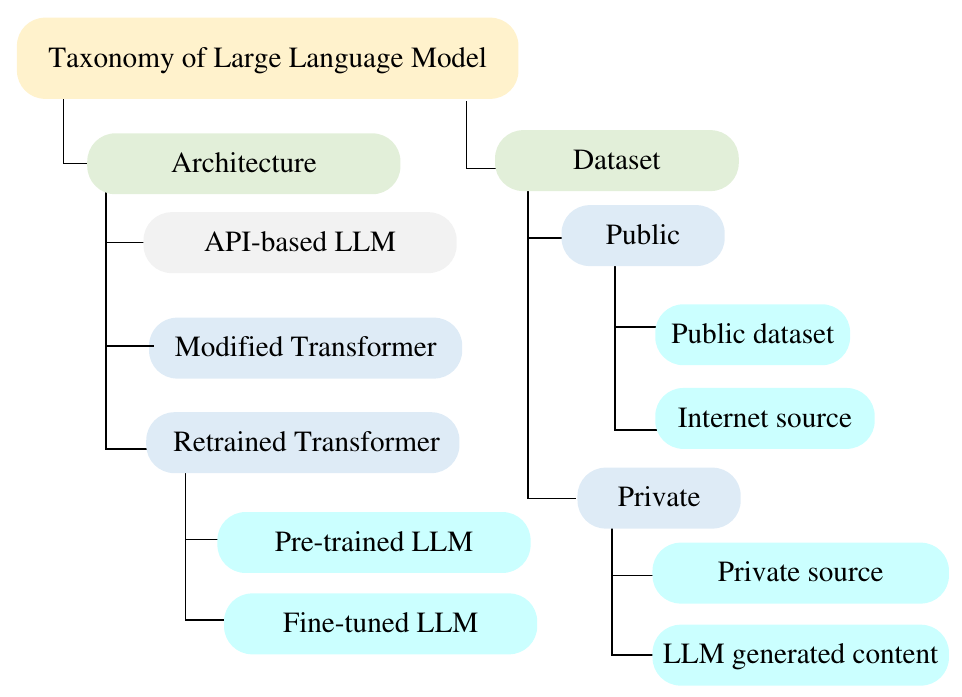}
\caption{Taxonomy of LLMs based on their architecture and dataset}
\label{fig:taxonomy}
\end{figure}

\noindent\textbf{P-(I) LLM Collection and Document Analysis:}
We collected a dataset comprising various LLMs, accompanied by comprehensive descriptions of their architectural frameworks and training datasets. These descriptions were meticulously sourced from official documentation, technical reports, and other authoritative resources provided by the creators of these models. To guarantee the accuracy and trustworthiness of our data, we conducted thorough manual reviews of the provided documentation and reports to confirm their origin from the model developers. Furthermore, to enhance the efficiency of our analysis, we employed an LLM to assist in the systematic examination of the technical reports and documentation. This strategy facilitated the precise extraction of critical information regarding the models' architectures and training methodologies.

After collecting all relevant documents, the next step involves categorizing the LLM based on the dataset and architecture, as shown in \autoref{fig:taxonomy}. Concerning architecture, we identify three primary types: Modified Transformer Architectures (e.g., Google’s BERT, which introduces bidirectional training, and OpenAI’s GPT-3, enhancing transformer scalability with new layers and modules), Pre-trained or Fine-tuned Models (such as RoBERTa, which adapts BERT through extended pre-training on a broader corpus), and API-Driven Models, where applications utilize the APIs of established models like IBM Watson without hosting the underlying architectures themselves. Regarding Training Data, we categorize it into Public Datasets, which are extensive and cover a wide range of topics suitable for general-purpose models (like the Common Crawl dataset used by OpenAI), and Private Datasets, which include data from private sources such as medical records collected from hospitals or synthesized datasets from models like GPT-3 for specific training enhancements, such as data augmentation to improve NLP task robustness.

% \subsubsection{Training Data}
% The training data for large language models can be categorized into two primary types:
% \begin{itemize}
% \item Internet-Derived and Public Datasets: The first category includes data sourced from the internet or publicly available datasets. These datasets are often vast in size and encompass a broad spectrum of topics, making them ideal for training general-purpose language models. 
% \item Data Generated by Other Large Models: The second category involves data that has been synthesized by other large language models.
% \end{itemize}

\vspace{2mm}
\noindent\textbf{P-(II) Self-Identity Recognition:}  The phase of self-identity recognition in LLMs is primarily concerned with evaluating the model's understanding and awareness of its own identity. This phase can be divided into three principal steps: (1) question collection, (2) question reformulation, and (3) response analysis. Each of these stages utilizes the LLM to further the investigation. Details on each of the steps are provided below: \looseness=-1

 \begin{itemize}
     \item \textbf{Question Collection.} During the question collection phase for assessing the self-identity cognition of LLMs, we prioritized sourcing a diverse array of questions from various platforms to ensure a comprehensive evaluation. These sources ranged from academic literature and online forums to expert interviews.
     \item \textbf{Question Reformulation.} Relying exclusively on test questions obtained from the internet may fail to provide a comprehensive evaluation framework. Moreover, it appears that some LLMs can adjust their self-identification through specialized fine-tuning processes. To mitigate these limitations and enhance test efficacy, we propose the use of LLMs to meticulously reformulate these gathered questions.  We specifically utilize rule-based techniques for this purpose, ensuring maximal test coverage by rewriting each sentence in all of the prescribed formats. For instance, when assessing a model originating from a country where English is the dominant language, we generate questions in Chinese, and vice versa.

    \item \textbf{Response Analysis.} We have re-engaged a LLM to facilitate the classification of responses. Specifically, the LLM was assigned to analyze responses from the models being tested and categorize them based on established criteria. This classification process is critical to our research on identity confusion, where we delineate six specific types. Correspondingly, we have developed numerous questions for each type, sourcing some from the internet and others generated by LLMs. The evaluation protocol is succinct: if a response from the LLM under examination matches any of the identified types of identity confusion, the assessment session is terminated. Conversely, if no such confusion is detected, we persist with the interrogation until we either classify the confusion type or confirm the absence of identity confusion in the model.
 \end{itemize}

\vspace{2mm}
\noindent\textbf{P-(III) Output Distribution Analysis:}
In this step, we assess the influence of structures or datasets on large models by evaluating similarities to pinpoint the origins of identity confusion. Output similarity has proven to be an effective approach for fingerprinting LLMs~\cite{yang2024fingerprint, russinovich2024hey, zhang2024reef}. When two LLMs with fundamentally distinct output distributions still demonstrate identity confusion, this strongly indicates that the confusion is unlikely to stem from one model directly using or replicating another. Instead, it is more plausibly attributed to hallucinations inherent in LLMs. 
%plagued by identity confusion likely share structural or dataset similarities. 
%These similarities could induce not only identity confusion but also similar outputs. 
Based on this hypothesis, our approach is as follows:
\begin{itemize}
    \item \textbf{Classification}: We categorize models with identity confusion into two groups—those with known structures and datasets, and those with unknown ones.
    \item \textbf{Correlation Analysis}: 
    Our experimental process consists of two key phases. In the first phase, we perform output testing on models sharing the same architecture and dataset (e.g., different versions of the same model from a single provider). If the results demonstrate significant output similarity, this would validate the effectiveness of output similarity as a reliable method for fingerprinting. In the second phase, we test the outputs of large models identified as exhibiting identity confusion. Should we find substantial differences in their outputs, it would suggest that the observed identity confusion is more likely a result of hallucinations inherent to large language models rather than direct replication or use of another model.

    %We gather outputs from models with known configurations and analyze their word distributions to determine whether output similarities are more pronounced among models sharing the same structure or dataset—or both. 
    
    %Our objective is to decipher whether the observed output similarities among models with identity confusion stem from structural or dataset similarities. We aim to establish a correlation between identity confusion and similarities in outputs, structures, and datasets.
    %\item \textbf{Application to Unknown Models}: In the final step, using the relationships we've identified, we analyze models with undisclosed structures and datasets. This analysis helps determine the probable causes of displayed identity confusion. For instance, if experiments show that two models with identity confusion and identical training datasets also exhibit similar outputs—despite having distinct structures—we can infer that similar outputs in other models with undisclosed information might hint at similar datasets, not necessarily similar structures.
\end{itemize}

 % Specifically, to that end, we initiate by preparing multiple datasets that encompass a variety of topics, styles, and complexities aimed at challenging the models and eliciting diverse outputs. We then collect the outputs from each dataset, analyzing their word distribution, which includes frequency metrics for different parts of speech and other linguistic features. To establish a baseline, we perform repeated tests on the same model to calculate the average distance between its runs, serving as a reference point for assessing the similarity or dissimilarity of output distributions across different models. We then employ similarity metrics such as cosine and Jaccard similarity to compare the word distribution similarity among the outputs from different models. Additionally, we use Euclidean distance to measure the straight-line distance between two points in a multidimensional space.
 
We prepare diverse datasets to challenge the models and elicit varied outputs. By analyzing word distributions, including part-of-speech frequencies, we establish a baseline through repeated tests on the same model to measure intra-model consistency. We then compare outputs across models using metrics like cosine and Jaccard similarity, along with Euclidean distance, to assess the similarity or dissimilarity in their output distributions.
 The Euclidean distance \( d \) between two points \(\mathbf{p} = (p_1, p_2, \ldots, p_n)\) and \(\mathbf{q} = (q_1, q_2, \ldots, q_n)\) in \( n \)-dimensional space is given by:
\begin{equation}
    d(\mathbf{p}, \mathbf{q}) = \sqrt{\sum_{i=1}^{n} (p_i - q_i)^2}
\end{equation}
 This represents the difference in part-of-speech distributions between two models given the same input, with a smaller Euclidean distance indicating closer output distributions and a higher similarity between models.

\subsection{Survey on Attitudes to Identity Confusion}
\label{subsec:survey}

To gain insights into the security implications stemming from identity confusion in LLMs, we utilized a structured survey approach, recruiting participants via Credamo. Our survey was meticulously designed to delve into users' understanding, usage patterns, and perceptions of LLM reliability. This approach enables us to comprehensively assess how such factors ultimately influence users' trust in LLMs.

\vspace{2mm}
\noindent\textbf{Design}: We segmented the survey into several distinct sections to ensure comprehensive coverage of the topic. The initial segment collected basic demographic information and general familiarity with LLMs. Subsequent sections were designed around specific scenarios depicting identity confusion instances involving LLMs. These scenarios were aimed to gauge trust levels and ascertain conditions under which users would still rely on LLMs. Please refer to the appendix for a detailed list of questions.To be more specific:
\begin{itemize}
    \item \textbf{Demographic Information}: This foundational section captured essential demographic details such as age, gender, location, and industry. This information helps contextualize subsequent responses based on the respondent’s background.

\item \textbf{General Familiarity with LLMs}: We gauged the participants' awareness and engagement with LLMs. Questions covered how often they use LLMs, how many they can name, and how many they have actively used. This section also included a question on whether they understood the basic working principles of LLMs.

\item \textbf{Identity Confusion Scenarios}: This part delved into specific, common scenarios involving LLMs that might affect user trust. Each scenario described a potential error or common misunderstanding (e.g., Self-Identification Confusion, Self-Identification Confusion) by using concrete example to ensure the user knows what we are referring to.  We explored user tolerance and trust in the technology when faced with such issues.
Respondents also required to rate their trust in LLMs on a five-point Likert scale. This scale was employed consistently across different scenarios to measure shifts in trust depending on the nature of the error or issue presented.

%\item \textbf{Open-ended Questions:} To supplement the structured Likert scale questions, we included open-ended questions where participants could elaborate on their reasons for trusting or distrusting LLMs, especially after encountering errors or misleading outputs.
\end{itemize}

\vspace{2mm}
\noindent\textbf{Participants}: To ensure the collection of high-quality data, we conducted a preliminary pilot survey targeting participants who met specific criteria, including a credit score greater than 80 and an approval rate higher than 90\%. These participants are recognized by Credamo for their reliability and attentiveness in completing tasks. Each participant in the main survey was compensated \$1 for their effort. From the overall responses, we systematically filtered out any that were rushed, illogical, or duplicated. This quality control step was crucial in ensuring that the responses we analyzed were only those that provided genuine insights into user attitudes towards LLMs.
Ultimately, our survey attracted 244 participants. After careful screening and removal of low-quality submissions, we compiled 208 valid responses for our analysis. This robust methodology allowed us to draw reliable conclusions about the general sentiment and trust levels users have towards large language models, enhancing our understanding of the public’s perception and acceptance of such technologies.

\section{Experiment}
\subsection{Experimental Setup}
\begin{table*}
\centering
\scriptsize
\setlength\tabcolsep{4pt}
\caption{Summary of LLMs with company, country, year, users, and technical specifications. We highlighted those with identity confusion. } 
\label{tab:summaryllm}

\begin{tabular}{llllllcccccc}
\toprule[1.5pt]
\multicolumn{1}{c}{\multirow{2}{*}{\textbf{Model}}} & \multicolumn{1}{c}{\multirow{2}{*}{\textbf{Company}}} & \multirow{2}{*}{\textbf{Country}} & \multirow{2}{*}{\textbf{Year}} & \multirow{2}{*}{\textbf{\# of Users}} & \multirow{2}{*}{\textbf{Parameters}} & \multirow{2}{*}{\textbf{OSS}} & \multicolumn{3}{c}{\textbf{Architecture}}                                                                                                                                              & \multicolumn{2}{c}{\textbf{Dataset}} \\ \cmidrule(lr){8-10} \cmidrule(lr){11-12} 
\multicolumn{1}{c}{}                                & \multicolumn{1}{c}{}                                  &                                   &                                &                                       &                                      &                               & \begin{tabular}[c]{@{}c@{}}Fine-tuned\\ LLM\end{tabular} & \begin{tabular}[c]{@{}c@{}}Pre-trained \\ LLM\end{tabular} & \begin{tabular}[c]{@{}c@{}}Modified\\ Transformer\end{tabular} & \textbf{Public}  & \textbf{Private}  \\ \hline\rowcolor[HTML]{FFCCC9} 
Baichuan2~\cite{baichuan}     & Baichuan                                              & China                             & 2023                           & Millions                              & 7B/13B                               & $\checkmark$                  & $\times$                                                 & $\checkmark$                                               & $\checkmark$                                                   & $\checkmark$     & $\checkmark$      \\
ChatGLM~\cite{chatglm}        & Zhipu                                                 & China                             & 2024                           & Millions                              & 9B                                   & $\checkmark$                  & $\times$                                                 & $\checkmark$                                               & $\checkmark$                                                   & $\checkmark$     & $\checkmark$      \\
GPT-3.5-Turbo~\cite{gpt3}     & OpenAI                                                & USA                               & 2023                           & Hundreds of millions                  & 175B                                 & $\times$                      & $\times$                                                 & $\checkmark$                                               & $\times$                                                       & $\checkmark$     & $\checkmark$      \\ \rowcolor[HTML]{FFCCC9} 
DeepSeek~\cite{DeepSeek}      & Deepseek                                              & China                             & 2023                           & Millions                              & 236B                                 & $\checkmark$                  & $\times$                                                 & $\checkmark$                                               & $\checkmark$                                                   & $\checkmark$     & $\times$          \\
Doubao~\cite{doubao}          & ByteDance                                             & China                             & 2023                           & Millions                              & -                                    & $\checkmark$                  & $\times$                                                 & $\checkmark$                                               & $\checkmark$                                                   & $\checkmark$     & $\times$          \\
ERNIE Bot~\cite{wenxin}       & BaiDu                                                 & China                             & 2023                           & Hundreds of millions                  & 260B                                 & $\times$                      & $\times$                                                 & $\checkmark$                                               & $\times$                                                       & $\checkmark$     & $\checkmark$      \\
Gemini~\cite{Gemini}          & Google                                                & USA                               & 2023                           & Hundreds of millions                  & 1560B                                & $\times$                      & $\times$                                                 & $\checkmark$                                               & $\times$                                                       & N/A              & N/A       \\
GPT2~\cite{gpt2}              & OpenAI                                                & USA                               & 2019                           & Tens of millions                      & 0.13B                                & $\checkmark$                  & $\times$                                                 & $\checkmark$                                               & $\checkmark$                                                   & $\checkmark$     & $\checkmark$      \\
GPT2-large                                          & OpenAI                                                & USA                               & 2019                           & Tens of millions                      & 0.81B                                & $\checkmark$                  & $\times$                                                 & $\checkmark$                                               & $\checkmark$                                                   & $\checkmark$     & $\checkmark$      \\
GPT2-xl                                             & OpenAI                                                & USA                               & 2019                           & Tens of millions                      & 1.61B                                & $\checkmark$                  & $\times$                                                 & $\checkmark$                                               & $\checkmark$                                                   & $\checkmark$     & $\checkmark$      \\
GPT4~\cite{GPT4}              & OpenAI                                                & USA                               & 2023                           & Hundreds of millions                  & 1800B                                & $\times$                      & $\times$                                                 & $\checkmark$                                               & $\times$                                                       & $\checkmark$     & $\checkmark$      \\
 Hunyuan~\cite{hunyuan}        & Tencent                                               & China                             & 2023                           & Millions                              & -                                    & $\times$                      & $\times$                                                 & $\checkmark$                                               & $\times$                                                       & N/A              & N/A      \\\rowcolor[HTML]{FFCCC9}
\rowcolor[HTML]{FFCCC9} 
Hailuo AI                     & Hailuo                                                & China                             & 2024                           & Millions                              & 15B                                  & $\checkmark$                  & $\times$                                                 & $\checkmark$                                               & $\checkmark$                                                   & $\checkmark$     & $\checkmark$      \\ 
Spark-Lite~\cite{spark}       & iFLYTEK                                               & China                             & 2023                           & Millions                              & -                                    & $\times$                      & $\times$                                                 & $\checkmark$                                               & $\times$                                                       & N/A              & N/A      \\
Spark-Pro                                           & iFLYTEK                                               & China                             & 2023                           & Millions                              & -                                    & $\times$                      & $\times$                                                 & $\checkmark$                                               & $\times$                                                       & N/A              &   N/A      \\
Spark-Max                                           & iFLYTEK                                               & China                             & 2023                           & Tens of millions                      & -                                    & $\times$                      & $\times$                                                 & $\checkmark$                                               & $\times$                                                       & N/A              & N/A      \\
Kimi~\cite{kimi}              & Moonshot                                              & China                             & 2023                           & Millions                              & 200B                                 & $\times$                      & $\times$                                                 & $\checkmark$                                               & $\times$                                                       & N/A              & N/A    \\
LLaMA2-egk                                          & Huggingface                                           & USA/France                        & 2023                           & Hundreds of thousands                 & 7B                                   & $\checkmark$                  & $\checkmark$                                             & $\times$                                                   & $\checkmark$                                                   & $\checkmark$     & $\checkmark$      \\
LLaMA2-hf                                           & Huggingface                                           & USA/France                        & 2023                           & Hundreds of thousands                 & 7B                                   & $\checkmark$                  & $\checkmark$                                             & $\times$                                                   & $\checkmark$                                                   & $\checkmark$     & $\checkmark$      \\
LLaMA2-wks                                          & Huggingface                                           & USA/France                        & 2023                           & Hundreds of thousands                 & 7B                                   & $\checkmark$                  & $\checkmark$                                             & $\times$                                                   & $\checkmark$                                                   & $\checkmark$     & $\checkmark$      \\
LLaMA2-ch~\cite{llama2}       & Meta                                                  & USA                               & 2023                           & Millions                              & 7B                                   & $\checkmark$                  & $\checkmark$                                             & $\times$                                                   & $\checkmark$                                                   & $\checkmark$     & $\checkmark$      \\ \rowcolor[HTML]{FFCCC9} 
MengZi~\cite{Mengzi}          & Langboat                                              & China                             & 2023                           & Millions                              & 8B/13B                               & $\checkmark$                  & $\times$                                                 & $\checkmark$                                               & $\checkmark$                                                   & $\checkmark$     & $\checkmark$      \\
%Qwen~\cite{qianwen}           & Alibaba                                               & China                             & 2023                           & Hundreds of millions                  & 7B/72B                               & $\checkmark$                  & $\times$                                                 & $\checkmark$                                               & $\checkmark$                                                   & $\checkmark$     & $\checkmark$      \\
Skywork~\cite{tiangong}       & Kunlun                                                & China                             & 2023                           & Millions                              & 13B                                  & $\checkmark$                  & $\times$                                                 & $\checkmark$                                               & $\checkmark$                                                   & N/A              & N/A      \\\rowcolor[HTML]{FFCCC9} 
Taichu~\cite{taichu}          & CAS                                                   & China                             & 2023                           & Millions                              & 32B                                  & $\times$                      & $\times$                                                 & $\checkmark$                                               & $\times$                                                       & N/A              & N/A     \\\rowcolor[HTML]{FFCCC9} 
Xiaobai~\cite{xiaobai}        & TianRang                                              & China                             & 2023                           & Hundreds of thousands                 & 186B                                 & $\times$                      & $\times$                                                 & $\checkmark$                                               & $\times$                                                       & N/A              & N/A       \\ \rowcolor[HTML]{FFCCC9} 
Yi~\cite{yi}                  & Yi                                                    & China                             & 2023                           & Millions                              & 7B/34B                               & $\checkmark$                  & $\times$                                                 & $\checkmark$                                               & $\checkmark$                                                   & N/A              & N/A       \\
360AI~\cite{360ai}            & 360                                                   & China                             & 2023                           & Millions                              & -                                    & $\times$                      & $\times$                                                 & $\checkmark$                                               & $\times$                                                       & N/A              & N/A     \\ \bottomrule[1.5pt]
\end{tabular}

  \label{tbl:table2}
\end{table*}

\noindent\textbf{LLMs.} As shown in \autoref{tab:summaryllm}, we assembled a comprehensive collection of 27 LLMs from various sources. These models were selected based on their diverse architectures, and training methodologies, ensuring a broad representation of the current state-of-the-art LLMs. %Among these models, a subset was also leveraged to assist in the implementation of our proposed pipeline, thereby facilitating an end-to-end evaluation process. 

\vspace{2mm}
\noindent\textbf{Questions Dataset.} For \textbf{RQ1}, we compiled identity confusion questions from internet platforms, refined them by removing duplicates, and used LLMs to expand the set, resulting in 77 unique questions for experiments. For \textbf{RQ2}, we utilized the HC3 dataset~\cite{hc3_dataset}, a benchmark comparing human-written and ChatGPT-generated text to evaluate quality and differences.

\subsection{Prevalence of Identity Confusion (RQ1)}
\label{subsec:q1}

Our evaluation encompassed a total of {27} distinct models. Among these, we identified that {7} models exhibited issues related to identity confusion, representing {25.93}\% of the entire set. This proportion underscores a significant vulnerability within a notable fraction of the models tested, highlighting a prevalent challenge in the design and training of LLMs to accurately manage and safeguard identity-related data. Interestingly, the previously controversial iFLYTEK Spark performed flawlessly in this test, successfully avoiding all identity confusion issues. Now, when any identity confusion-related question is input, its response is consistently: ``\textit{Hello, I am iFLYTEK Spark, developed by iFLYTEK, and my name is iFLYTEK Spark. I can communicate with human beings naturally, answer questions, and efficiently fulfill the needs of cognitive intelligence across various fields.''} The overall experimental results effectively address \textbf{RQ1}. \looseness=-1

\begin{mdframed}[backgroundcolor=blue!4]  
\noindent\textit{\textbf{Finding~(I).}{ 
Previously vulnerable models can often be improved rapidly. For instance, iFLYTEK Spark (previously vulnerable) demonstrated remarkable progress, flawlessly handling identity-related queries and delivering consistent, accurate responses.}}
\end{mdframed}

\vspace{2mm}
\noindent\textbf{Intellectual Property.} 
Our analysis distinguished between open-source and proprietary models. Among the open-source LLMs, {5} out of {15} ({30}\%) exhibited identity confusion. In contrast, the proprietary models showed a slightly higher incidence, with {2} out of {12} ({17}\%) affected. This suggests that while open-source models are generally more accessible for scrutiny and potentially more susceptible to identity-related vulnerabilities, the proprietary models also share a considerable risk, albeit slightly lower. This could indicate that access to source code alone does not necessarily correlate with higher security in terms of identity confusion.

\vspace{2mm}
\noindent\textbf{Architecture.} 
From the architectural perspective, we classified the models into three categories: pre-trained LLMs, modified Transformer architectures, and fine-tuned LLMs. As shown in \autoref{tab:ic}, the findings revealed that 7 out of 23 (30\%) pre-trained LLMs faced identity confusion issues. Modified Transformers showed 5 out of 15 (30\%) models affected.  Pre-trained models and Modified Transformers are more vulnerable because of their broad training scope, while fine-tuning  refines the model's ability to handle ambiguities, leading to a decrease in susceptibility.

\vspace{2mm}
\noindent\textbf{Training Dataset.} 
In our study, we also investigated the impact of dataset types—public versus private—on the incidence of identity confusion within LLMs. However, detailed information about data sources is often not disclosed by many LLMs. As a result, our analysis is based on the data that was publicly available. Through document analysis, we identified that 16 models incorporate public datasets, while 14 utilize private datasets in their training processes.  As shown in \autoref{tab:ic}, our analysis showed that LLMs trained on public datasets encountered identity confusion in 4 out of 16 (25\%), whereas those utilizing data from private datasets experienced a higher rate of 3 out of 14 (21\%).  
This suggests that models trained on less curated internet-sourced data are more susceptible to identity confusion, potentially due to the unstructured and diverse nature of online information which might contain more ambiguous and conflicting identity cues.
% \subsubsection{Question Reformulation}
% Through the application of techniques such as paraphrasing and expansion to modify our test questions, we have observed suspicious responses from certain models. 

\begin{figure}[ht]
  \centering
  \includegraphics[width=0.5\textwidth]{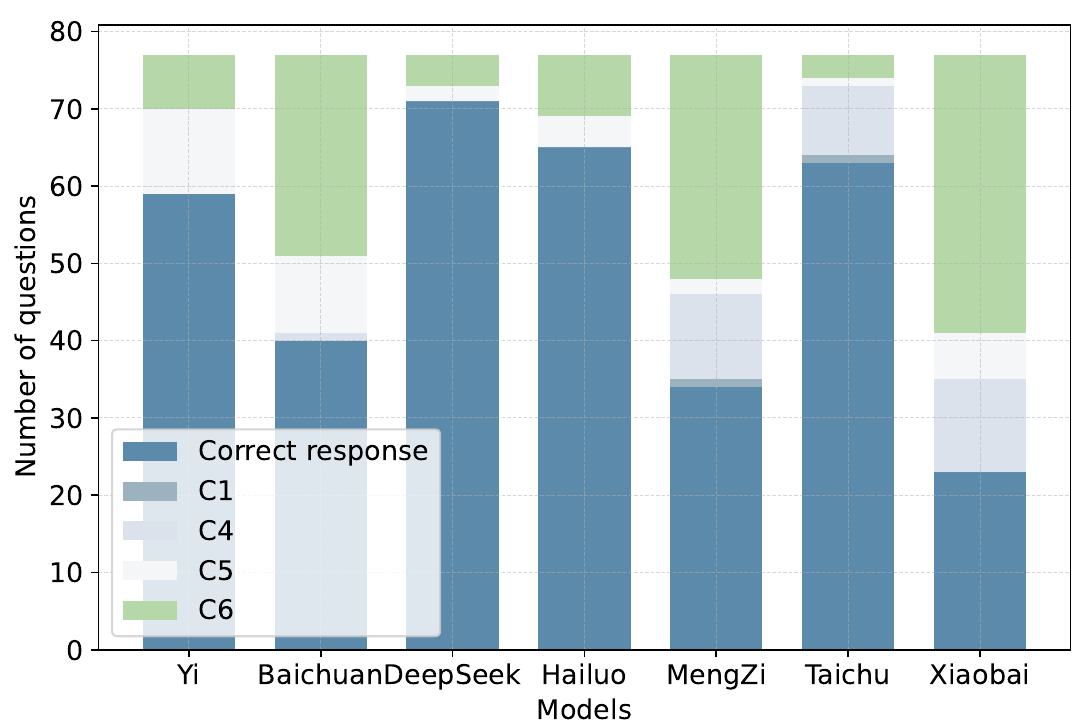}
  \caption{Types of identity confusion in our experiment.}
  \label{fig:fig4}
\end{figure}

\vspace{2mm}
\noindent\textbf{Types of Identity Confusion.}
In the analysis of various types of identity confusion within LLMs, our study categorized confusion into six distinct types: self-identification Confusion, reference confusion, capabilities confusion, profile confusion, relationship confusion and creation confusion. As shown in \autoref{fig:fig4}, we excluded types of identity confusion that were not identified in the experiment. In total, we identified four distinct types of identity confusion, with creation confusion emerging as the most prevalent, accounting for 63.13\% of all observed identity confusion issues. 
Profile confusion and relationship confusion were also widely observed across the models. Surprisingly, self-identity confusion and reference confusion, which have been extensively reported in previous studies, were almost absent in our experiments. We infer that the widespread attention to these issues has prompted vendors to address and resolve them effectively. Interestingly, we also observed that if an LLM exhibits one type of identity confusion, it is likely to exhibit other types as well.
These results underscore the complexity of designing LLMs that can accurately interpret and represent various aspects of identity, especially in dynamically evolving contexts where entity roles and relationships can be nuanced and multifaceted.  
% \usepackage{booktabs}
% \usepackage{multirow}
\begin{table}[]
\scriptsize

\setlength\tabcolsep{5pt}
 \caption{Identity Confusion w.r.t. IP, architecture, dataset}
  \label{tab:ic}
\begin{tabular}{@{}clccc@{}}
\toprule
\multicolumn{2}{c}{\textbf{Category}}                                        & \multicolumn{1}{l}{\textbf{Total}} & \textbf{\begin{tabular}[c]{@{}c@{}}\# of Identity Confusion \end{tabular}} & \textbf{\begin{tabular}[c]{@{}c@{}} \textbf{\%} \end{tabular}} \\ \midrule

\multirow{2}{*}{\textbf{\begin{tabular}[l]{@{}c@{}}IP \end{tabular}}} & \textbf{Open-source} & 15  & 5   & 30  \\ & \textbf{Proprietary}           & 12 & 3  & 25   \\ \midrule

\multirow{3}{*}{\textbf{Architecture}}  & \textbf{Pre-trained LLMs}      & 23 & 7  & 30   \\ & \textbf{Modified Transformers} & 15   & 5   & 30  \\ & \textbf{Fine-tuned LLMs} & 4 & 0  & 0   \\ \midrule

\multirow{2}{*}{\textbf{Dataset}} & \textbf{Public Dataset}  & 16  & 4  & 25 \\ & \textbf{Private Dataset} & 14   & 3    & 21  \\ \bottomrule  
\end{tabular}

\end{table}

\begin{mdframed}[backgroundcolor=blue!4]  
\noindent\textit{\textbf{Finding~(II).}{ 
Creation confusion emerges as the most prevalent form of identity confusion, constituting 63.13\% of all observed instances. Furthermore, the presence of one type of identity confusion in an LLM strongly correlates with the likelihood of exhibiting other types.}}
\end{mdframed}

\begin{figure}[htp]
    \centering
    \includegraphics[width=0.55\textwidth]{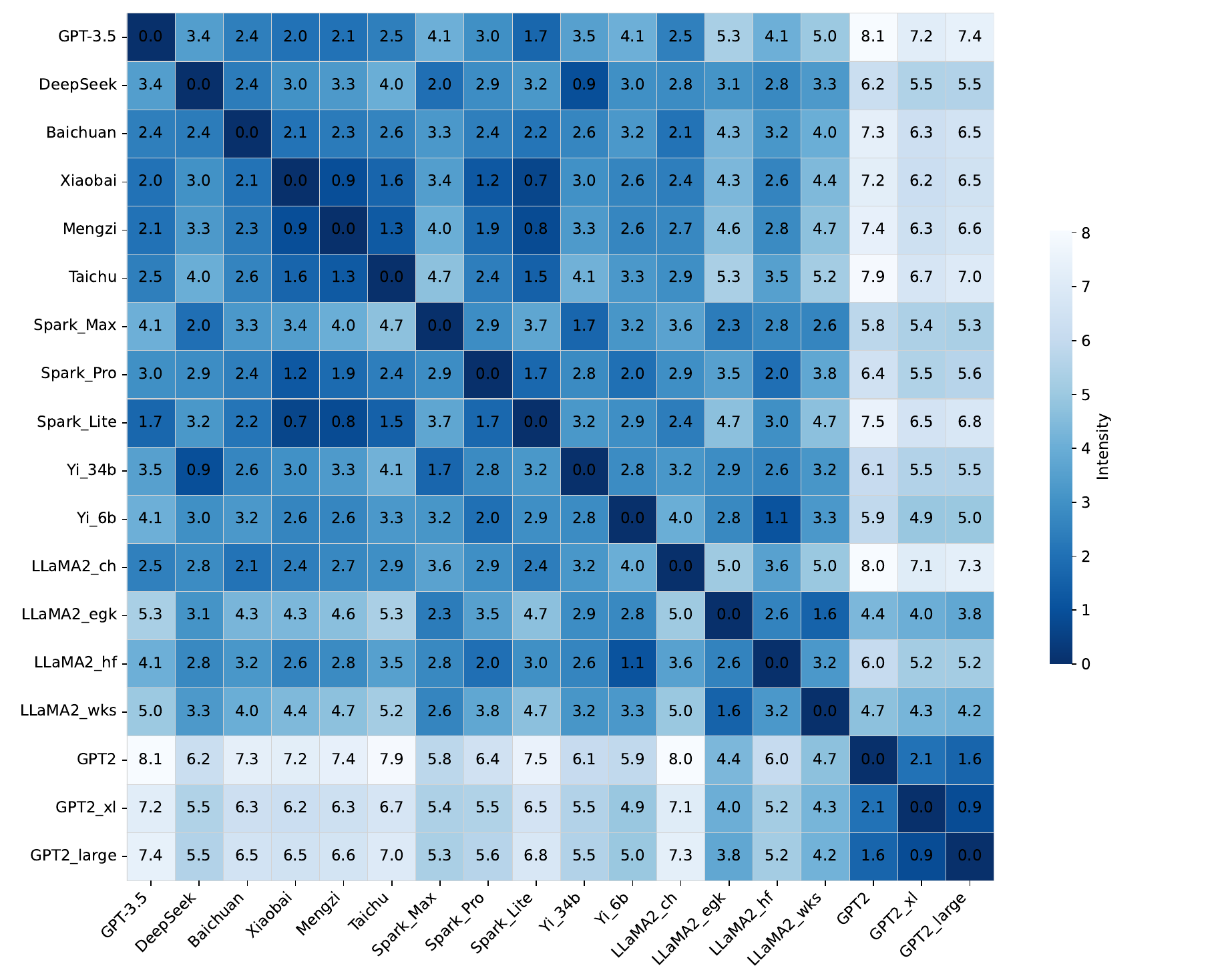}
    \caption{The heatmap illustrates the output similarity across different LLMs, with darker colors indicating higher levels of similarity. Please note that Hailuo AI was excluded from our evaluation due to the lack of an API for testing.}
    \label{fig:fig5}
\end{figure}

\begin{figure}[htbp]
    \centering
    \begin{minipage}{0.24\textwidth} % Adjust width as needed
        \centering
        \includegraphics[width=\textwidth]{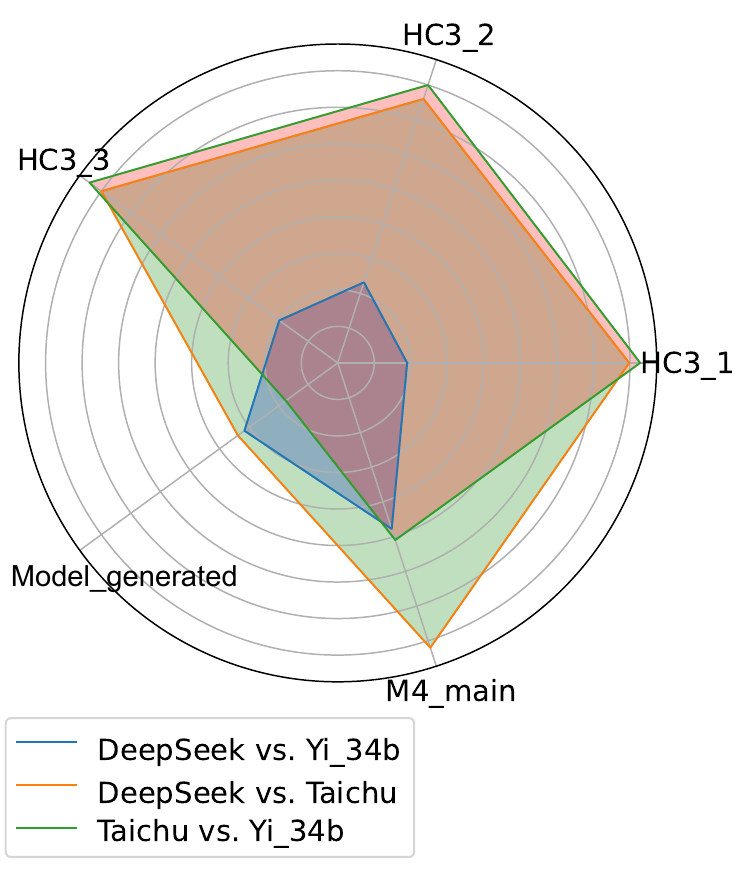} 
       
        \label{fig:fig7}
    \end{minipage}%
    \begin{minipage}{0.24\textwidth} % Adjust width as needed
        \centering
        \includegraphics[width=\textwidth]{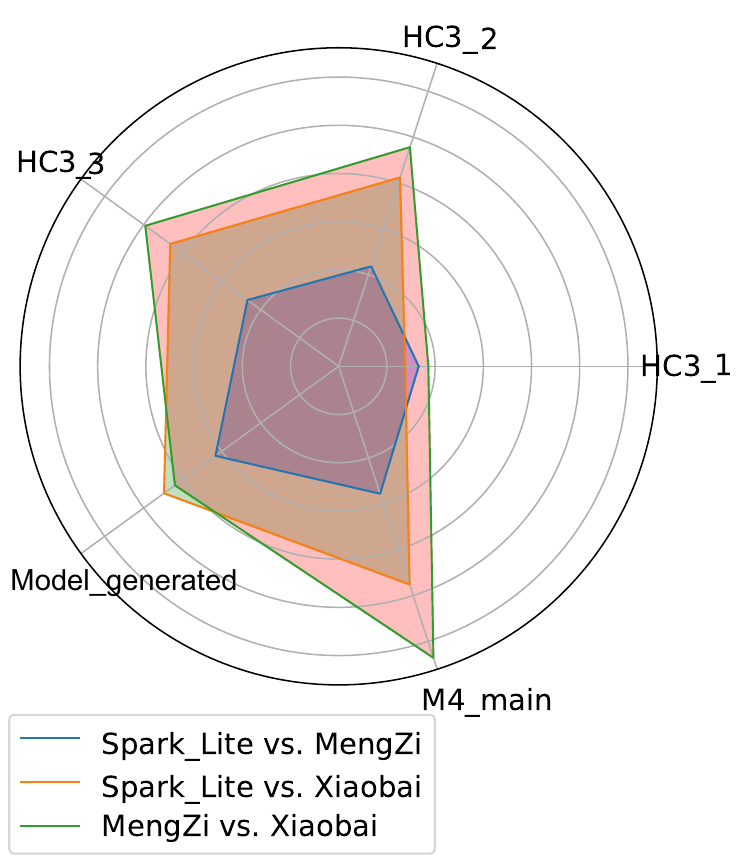}
        
        \label{fig:fig8}
    \end{minipage}
\caption{Radar chart of model pair output similarities across dataset subsets}
\label{fig:7}
\end{figure}

\subsection{Cause Analysis of Identity Confusion (RQ2)}

As shown in the \autoref{tab:summaryllm}, six LLMs exhibit identity confusion. To evaluate the output similarity of these models, we employ POS analysis. Specifically, we measure the outputs of both the model under evaluation and the model it is confused with (e.g., if Yi is confused with ChatGPT, we analyze the outputs of both Yi and ChatGPT). Additionally, different versions of the models are included to serve as benchmarks.  The analysis revealed high output similarity between different versions of the same model. For example, when comparing GPT2-xl and GPT2-large, we noted a low Euclidean distance of {0.9} across the word distribution data.  This outcome suggests that models within the same series exhibit a high degree of similarity in their outputs (e.g., the GPT2 series, Spark-Lite and Spark-Pro). This also corroborates the conclusions~\cite{fingerprinting}.

However, we observed that some models developed by different companies also exhibit relatively high similarity in their output distributions (e.g., DeepSeek and Yi-34b, as well as Xiaobai, Mengzi, and Spark-Lite).
To deepen our understanding of this similarity, we conducted additional experiments, running multiple rounds of testing with the same models on an identical subset of the HC3 dataset. The results are presented in \autoref{fig:fig7}.
The radar chart compares the output similarities of model pairs (e.g., DeepSeek vs. Yi\_34b, DeepSeek vs. Taichu) across different dataset subsets. Each axis represents a subset or feature, and the colored areas indicate similarity scores. Larger or overlapping areas suggest higher similarity, while smaller or irregular areas indicate variability.
Our experimental results reveal that while two models may exhibit similar output distributions in \autoref{fig:fig5}, their distributions can differ significantly when tested on a different dataset, as shown in \autoref{fig:fig7}. This allows us to answer RQ2: Different LLMs generally have distinct output distributions, while the same LLMs maintain similar distributions. Notably, LLMs with identity confusion issues exhibited significantly divergent output distributions, confirming they are distinct models.

\begin{mdframed}[backgroundcolor=blue!4]  
\noindent\textit{\textbf{Finding~(III).}{ 
Different LLMs typically exhibit distinct output distributions, whereas variations or versions of the same LLM maintain consistent distributions. Therefore, since the LLMs displaying identity confusion showed significant divergence in their output distributions, they can be identified as unique LLMs.}}
\end{mdframed}

\subsection{Implications of Identity Confusion (RQ3)}

\begin{figure}[htp]
    \centering
    \includegraphics[width=0.45\textwidth]{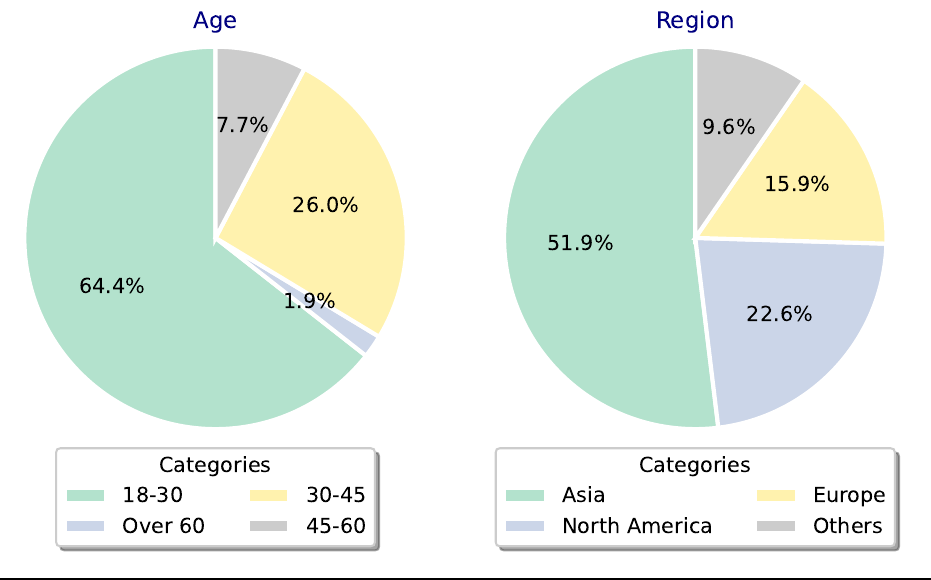}
    \caption{Age and region distribution of LLM users}
    \label{fig:age}
\end{figure}

\begin{figure}[htp]

    \centering
    \includegraphics[width=0.45\textwidth]{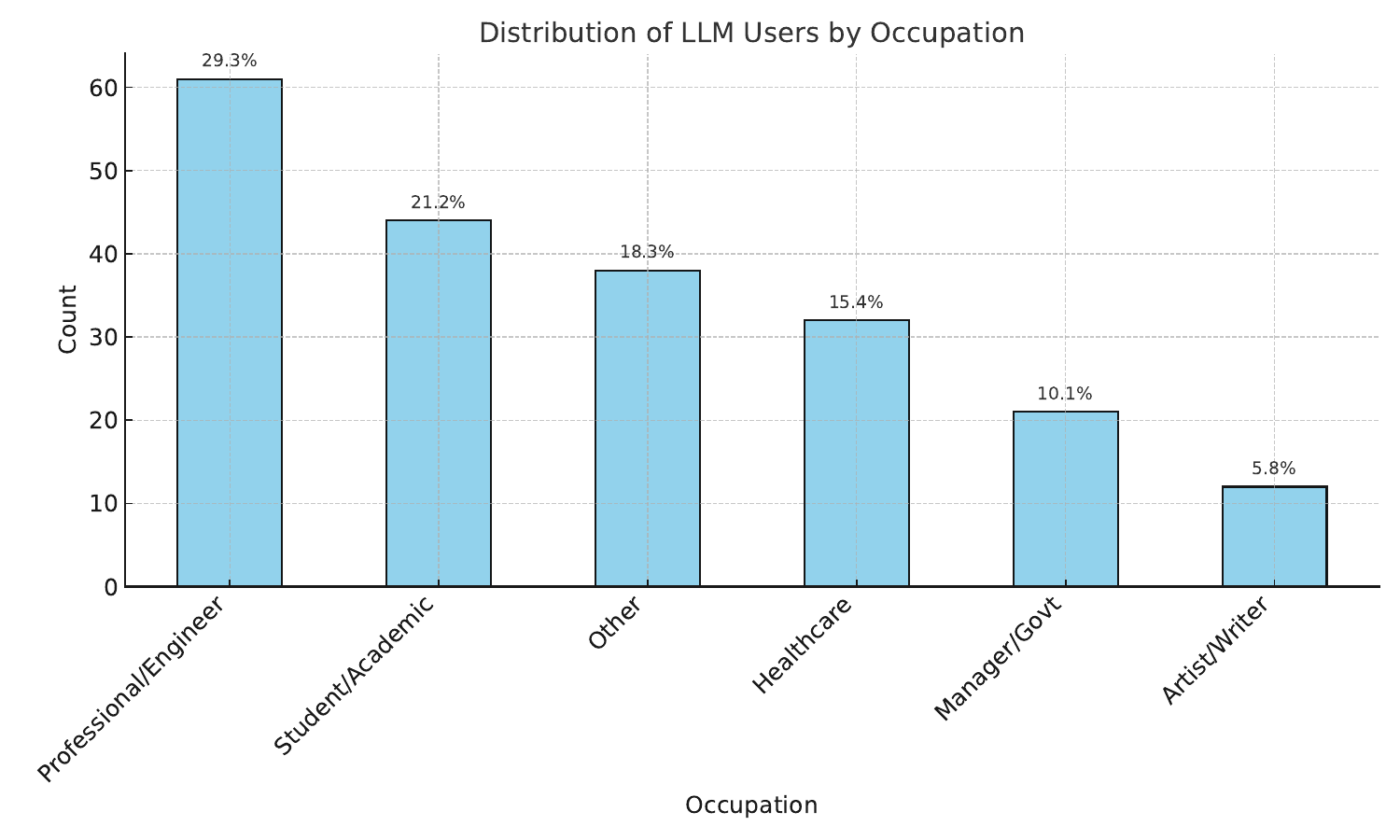}
    \caption{Occupation distribution of LLM Users}
    \label{fig:occupation}
\end{figure}

\noindent\textbf{Demographic Overview}:
To ensure the accuracy and reliability of our experimental results, our survey encompassed a broad demographic spectrum, incorporating various age groups, genders, occupations and geographical regions. 
Our survey is exclusively for individuals aged 18 and older who have experience using LLMs.
As shown in \autoref{fig:age}, the sample consisted of 208 participants, including 94 males and 107 females, with 7 individuals preferring not to disclose their gender. This distribution yields a nearly 1:1 male-to-female ratio, enhancing the representativeness of our study.
The majority (51.9\%) are from Asia, signifying that this region dominates the dataset.
North America follows with 22.6\%, while Europe accounts for 15.9\%. 
The dominance of Asia and North America in LLM usage can be attributed to the rapid growth of LLM development in countries like India, China, and the United States. The majority of people using LLMs fall within the 18-30 age range (64.4\%) and 31-45 age range (26.0\%), and older users (45-60 and over 60 combined) represent only a small portion of the user base.  
This could suggest that LLM technology is particularly appealing or relevant to younger demographics, possibly due to factors like higher digital literacy. As shown in \autoref{fig:occupation}, professional/Engineer and Student/Academic categories dominate the distribution, indicating that LLMs are highly utilized in technical, educational, and academic fields. Also,
The significant representation of the Healthcare category suggests that LLMs are being used for tasks such as medical documentation, diagnostics, or patient communication, highlighting their growing role in the healthcare sector. \looseness=-1
%Q1-2

\begin{figure}[htp]
    \centering
    \includegraphics[width=0.45\textwidth]{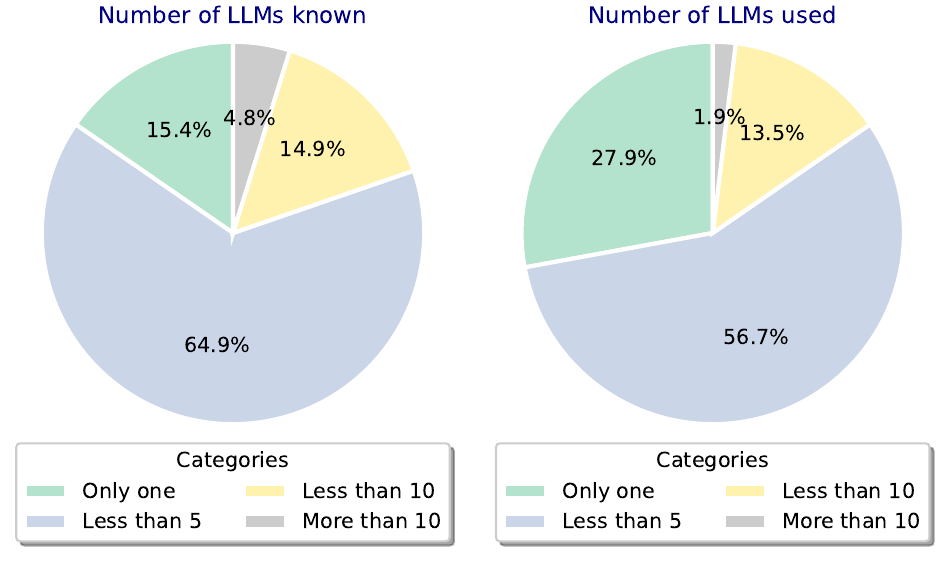}
    \caption{Familiarity distribution of LLM users}
    \label{fig:llmf}
\end{figure}

\begin{table*}[]
\footnotesize
 
\setlength\tabcolsep{1pt}
    \centering
   
    \begin{tabular}{lrrrrrrrrrrrrrrrrrrrrrrrr}
\toprule[1.5pt] 
\multirow{3}{*}{\textbf{Task}}
&\multirow{3}{*}{\textbf{Initial}}
&\multicolumn{14}{c}{{\bf Circumstance}}\\
\cmidrule(lr){3-18} 
&\multicolumn{1}{c}{{\textbf{}}}
&\multicolumn{2}{c}{{\textbf{Fault I}}}
&\multicolumn{2}{c}{{\textbf{Fault II}}}
&\multicolumn{2}{c}{{\textbf{C1}}}
&\multicolumn{2}{c}{\textbf{{C2}}}
&\multicolumn{2}{c}{{\textbf{C3}}}
&\multicolumn{2}{c}{{\textbf{C4}}}
&\multicolumn{2}{c}{{\textbf{C5}}}
&\multicolumn{2}{c}{{\textbf{C6}}}\\
\cmidrule(lr){3-4}\cmidrule(lr){5-6} \cmidrule(lr){7-8} \cmidrule(lr){9-10} \cmidrule(lr){11-12} \cmidrule(lr){13-14} \cmidrule(lr){15-16}  \cmidrule(lr){17-18}

&&Num.&$\Delta$&Num.&$\Delta$&Num.&$\Delta$&Num.&$\Delta$&Num.&$\Delta$&Num.&$\Delta$&Num.&$\Delta$ &Num.&$\Delta$
\\ \midrule

Writing Assistance & 120
&102&\textcolor{red}{-15.0\%$\downarrow$}
& 80&\textcolor{red}{-33.3\%$\downarrow$}  
& 74&\textcolor{red}{-38.3\%$\downarrow$} 
& 68&\textcolor{red}{-43.3\%$\downarrow$}
& 71&\textcolor{red}{-40.8\%$\downarrow$}
& 64&\textcolor{red}{-46.7\%$\downarrow$}
& 64&\textcolor{red}{-46.7\%$\downarrow$}
& 64&\textcolor{red}{-46.7\%$\downarrow$}
\\

Creative writing & 102
&101&\textcolor{red}{-1.0\%$\downarrow$}
&102&{0.0\%}  
&70&\textcolor{red}{-31.4\%$\downarrow$} 
&73&\textcolor{red}{-28.4\%$\downarrow$} 
&68&\textcolor{red}{-33.3\%$\downarrow$}
&72&\textcolor{red}{-29.4\%$\downarrow$}
&72&\textcolor{red}{-29.4\%$\downarrow$}
&73&\textcolor{red}{-28.4\%$\downarrow$}
\\
General Q\&A & 92
&91&\textcolor{red}{-1.1\%$\downarrow$}
&68&\textcolor{red}{-26.1\%$\downarrow$}  
&69&\textcolor{red}{-25.0\%$\downarrow$} 
&55&\textcolor{red}{-40.2\%$\downarrow$} 
&61&\textcolor{red}{-33.7\%$\downarrow$}
&68&\textcolor{red}{-26.1\%$\downarrow$}
&58&\textcolor{red}{-37.0\%$\downarrow$}
&61&\textcolor{red}{-33.7\%$\downarrow$}
\\ 
Professional Q\&A & 38
&25&\textcolor{red}{-34.2\%$\downarrow$}
&21&\textcolor{red}{-44.7\%$\downarrow$}  
&18&\textcolor{red}{-52.6\%$\downarrow$} 
&15&\textcolor{red}{-60.5\%$\downarrow$} 
&23&\textcolor{red}{-39.5\%$\downarrow$}
&15&\textcolor{red}{-60.5\%$\downarrow$}
&15&\textcolor{red}{-60.5\%$\downarrow$}
&22&\textcolor{red}{-42.1\%$\downarrow$}
\\
Educational purposes & 113
&68&\textcolor{red}{-39.8\%$\downarrow$}
&56&\textcolor{red}{-50.4\%$\downarrow$}  
&43&\textcolor{red}{-61.9\%$\downarrow$} 
&40&\textcolor{red}{-64.6\%$\downarrow$} 
&40&\textcolor{red}{-64.6\%$\downarrow$}
&39&\textcolor{red}{-65.5\%$\downarrow$}
&37&\textcolor{red}{-67.3\%$\downarrow$}
&40&\textcolor{red}{-64.6\%$\downarrow$}
\\
Personal entertainment & 65
&69&\textcolor{blue}{+6.2\%$\uparrow$}
&60&\textcolor{red}{-7.7\%$\downarrow$}  
&72&\textcolor{blue}{+10.8\%$\uparrow$} 
&70&\textcolor{blue}{+7.7\%$\uparrow$} 
&85&\textcolor{blue}{+30.8\%$\uparrow$}
&77&\textcolor{blue}{+18.5\%$\uparrow$}
&84&\textcolor{blue}{+29.2\%$\uparrow$}
&84&\textcolor{blue}{+29.2\%$\uparrow$}
\\
Professional tasks & 61
&54&\textcolor{red}{-11.5\%$\downarrow$}
&37&\textcolor{red}{-39.3\%$\downarrow$}  
&30&\textcolor{red}{-50.8\%$\downarrow$} 
&30&\textcolor{red}{-37.7\%$\downarrow$} 
&39&\textcolor{red}{-36.1\%$\downarrow$}
&32&\textcolor{red}{-47.5\%$\downarrow$}
&40&\textcolor{red}{-34.4\%$\downarrow$}
&37&\textcolor{red}{-39.3\%$\downarrow$} 
\\ 

\midrule
 
TOTAL&591
&510&\textcolor{red}{-13.7\%$\downarrow$}
&424&\textcolor{red}{-28.3\%$\downarrow$}
&376&\textcolor{red}{-36.4\%$\downarrow$}
&351&\textcolor{red}{-40.6\%$\downarrow$}
&387&\textcolor{red}{-34.5\%$\downarrow$}
&367&\textcolor{red}{-37.9\%$\downarrow$}
&370&\textcolor{red}{-37.4\%$\downarrow$}
&381&\textcolor{red}{-35.5\%$\downarrow$}
\\

\bottomrule[1.5pt]
\\
 \end{tabular}
   \caption{In different scenarios, the changes in the number of respondents willing to continue using LLMs for various tasks. C1 to C6 represent six types of identity confusion.}  
    \label{tab:vuln_cat_full}

\end{table*}

\vspace{2mm}

\vspace{2mm}
\noindent\textbf{User's Familiarity with LLMs}: 
To assess participants' familiarity with LLMs, we collected data on the number of LLMs they have used and are familiar with.  \autoref{fig:llmf} reveals that most users are familiar with fewer than 5 LLMs (64.9\%), and actual usage is similarly limited, with 56.7\% having used fewer than 5 models. A significant portion relies on only one LLM (15.4\% for familiarity, 27.9\% for usage). Few users explore a broader range, with just 4.8\% familiar with more than 10 and 1.9\% having used more than 10. To know the understanding of LLMs for those user, 
we utilized a 5-point  scale, where 1 represents ``No understanding'' and 5 signifies ``Thorough understanding''. 
To provide users with a clear understanding of the defined familiarity levels, we offer examples for each category. For instance, a ``thorough understanding'' is exemplified by the statement, ``\textit{I am actively involved in developing LLMs and have a comprehensive knowledge of how they work.}'' In contrast, ``no understanding'' is characterized by the statement, ``\textit{I have never explored this topic or only possess a superficial awareness.}'' Similarly, ``very little understanding'' is described as, ``\textit{I have heard about it from others and can vaguely recall some concepts, but I do not understand how it functions}.''
Interestingly, A majority (nearly 66\%) of users fall within the moderate-to-general understanding range, highlighting a broad but non-expert user base. A smaller group (14.5\%) represents developers and experts with a deep understanding, reflecting LLMs' niche technical complexity. \looseness=-1

\vspace{2mm}
\noindent\textbf{Usage of LLMs}:
We also examined the purposes for which LLMs are used. The findings reveal that LLMs are predominantly utilized for writing assistance (72.3\%, 86 respondents), educational purposes (68.1\%, 81 respondents), and general Q\&A (61.3\%, 73 respondents), underscoring their significant role in professional, academic, and everyday tasks. Creative applications, such as generating stories and ideas, also garnered substantial interest (53.8\%, 64 respondents), highlighting their appeal for imaginative use cases.
Professional tasks, including code generation and translation, represent a smaller but meaningful segment of usage (36.9\%, 44 respondents). Meanwhile, personal entertainment (32.8\%, 39 respondents) and professional Q\&A (29.4\%, 35 respondents) are less common. Only 1.7\% (2 respondents) reported ``Other'' purposes, with one using LLMs for job hunting and the other for conducting research.

\vspace{2mm}
\noindent\textbf{User Trust in LLMs When Facing Identity Confusion}:
We assessed users' trust in LLMs when they encountered identity confusion. Participants were presented with real-life scenarios, accompanied by examples, representing different types of identity confusion (C1–C6). They were then asked whether they would still use the LLM for the task in question.
The ``Initial'' value represents the number of respondents before encountering identity confusion, while the ``Num.'' under each circumstance (e.g., C1–C6) reflects the adjusted number of respondents after encountering a specific type of identity confusion. A positive value indicates an increase in trust, whereas a negative value signifies a decline. \looseness=-1

As shown in \autoref{tab:vuln_cat_full}, we can observe that user tolerance for identity confusion in LLMs varies significantly depending on the task. Critical tasks, such as Educational Purposes, Professional Q\&A, and Writing Assistance, show the sharpest declines in user willingness when faced with identity confusion. For example, Educational Purposes experienced the largest decrease (-67.3\% in C5, Relationship Confusion), highlighting the importance of trust and accuracy in tasks where factual correctness is paramount. Similarly, Professional Q\&A and Writing Assistance saw significant declines across multiple types of identity confusion, underscoring that users demand consistency and reliability in professional and high-stakes contexts. Interestingly, n contrast, less critical tasks, such as Personal Entertainment and Creative Writing, demonstrate higher resilience to identity confusion. Personal Entertainment even shows positive trends in some scenarios, with an increase of +10.8\% for C1 (Self-Identification Confusion) and +30.8\% for C3 (Capabilities Confusion). This is understandable. For instance, imagine the LLM saying, \textit{``I am ChatGPT,''} when it is actually DeepSeek. The user might find this amusing rather than concerning. \looseness=-1

\begin{mdframed}[backgroundcolor=blue!4]  
\noindent\textit{\textbf{Finding~(IV).}{ The findings show that user tolerance for identity confusion in LLMs depends on the task. Critical tasks (e.g., Educational Purposes) see sharp declines in user willingness, reflecting the need for trust and accuracy. In contrast, less critical tasks such as Personal Entertainment  show higher resilience.
 }}
\end{mdframed}

Certain types of identity confusion have more significant negative impacts as they directly undermine core elements of user trust, reliability, and the perceived competence of LLMs. For example, among these identity confusion, reference confusion (C2) consistently stands out as the most disruptive across tasks, showing the largest decline in user trust. This may be because this type of confusion has an immediate and direct impact—for instance, providing incorrect references for development documents. Such errors not only affect the specific task but also leave users with the impression that the LLM is unreliable across other tasks as well. \looseness=-1

\begin{mdframed}[backgroundcolor=blue!4]  
\noindent\textit{\textbf{Finding~(V).}{Certain types of identity confusion have more significant negative impacts. Reference confusion (C2) has the most significant negative impact, as it directly undermines user trust and reliability.  
 }}
\end{mdframed}

\begin{table}
\centering
\scriptsize

\setlength\tabcolsep{2pt}
\begin{tabular}{lrr}
\toprule[1.5pt]
            \textbf{Attitude Toward LLM Identity Confusion} &  \#  &    (\%) \\
\midrule
It makes me disappointed with the company that developed the model & 102 & 49.04 \\
It has no impact on my attitude & 61 & 29.33 \\
It makes me disappointed with artificial intelligence in general & 41 & 19.71 \\
Other  & 4 & 1.92 \\ \midrule
          \textbf{ Perceived Causes of LLM Identity Confusion} &  \#  &    (\%) \\
\midrule
 learned incorrect identity information from training data &                     74 &           35.58 \\
There are errors in the model's design or operation &                     71 &           34.13 \\
The model may have plagiarized other models &                     62 &           29.81 \\
Other  &                      1 &            0.48 \\
 
\bottomrule[1.5pt]
\end{tabular}
\vspace{2mm}
\caption{Impact of Identity Confusion on User Attitudes}
\label{tab:identity_confusion}
\end{table}

\vspace{2mm}
\noindent\textbf{Comparison of Identity Confusion with Other Faults}:
The experiment examined how users' willingness to continue using LLMs varied across different tasks and scenarios, focusing on Fault I (logical or numerical errors), Fault II (response inconsistencies), and six types of identity confusion (C1–C6). Participants were presented with two questions, each representing a specific fault:
\begin{itemize}
    \item [\textbf{Q1:}] If a large language model states that 3.11 is greater than 3.8, would you still trust the other answers provided by the model? (Fault I)
    \item [\textbf{Q2:}] If a large language model provides inconsistent answers when asked the same question twice, would you still trust its other answers?  (Fault II)
\end{itemize}
\noindent These issues were selected because they are widely recognized and frequently reported by users. As shown in \autoref{tab:vuln_cat_full},  Fault I and Fault II had moderate, task-specific impacts, with overall declines of 13.7\% and 28.3\%, respectively, but were generally seen as fixable technical problems. In contrast, identity confusion significantly undermined trust, with user willingness dropping sharply across all cases (each exceeding 35\%).

%In contrast, identity confusion undermined trust and evoked broader dissatisfaction, with 49.04\% of respondents blaming the company and 19.71\% expressing disappointment with AI in general. The findings emphasize that identity confusion poses a systemic threat to LLM adoption, making improvements in transparency, consistency, and contextual awareness crucial to rebuilding trust and maintaining user engagement.

\begin{mdframed}[backgroundcolor=blue!4]  
\noindent\textit{\textbf{Finding~(VI).}{The experiment showed that while logical errors and response inconsistencies had moderate, fixable impacts on user trust, identity confusion posed a systemic threat, significantly undermining trust. 
 }}
\end{mdframed}
% Tasks ranged from Writing Assistance and Educational Purposes to Personal Entertainment and Professional Q\&A. Results showed that identity confusion (e.g., C4: Profile Confusion, C5: Relationship Confusion) consistently caused the largest declines in user willingness across all tasks, with critical applications like Educational Purposes (-67.3%) and Professional Q\&A (-60.5%) being the most affected. Fault I and Fault II had moderate, task-specific impacts, with declines of up to -39.8% and -50.4%, respectively, but were generally perceived as fixable technical issues. In contrast, identity confusion undermined trust and evoked broader dissatisfaction, with 49.04% of respondents blaming the company and 19.71% expressing disappointment with AI in general. The findings emphasize that identity confusion poses a systemic threat to LLM adoption, making improvements in transparency, consistency, and contextual awareness crucial to rebuilding trust and maintaining user engagement.

\vspace{2mm}
\noindent\textbf{Attitude and Perceived Cause:} 
We also collected  the impact of identity confusion in LLMs on user attitudes and their perceived causes of such confusion.  
As shown in \autoref{tab:identity_confusion},  102 respondents expressed disappointment with the companies developing large language models due to identity confusion, representing 49.0\% of the total participants. Additionally, 61 individuals, accounting for 29.3\% of the respondents, reported no change in their attitudes towards these companies. Furthermore, 41 participants, or 19.7\% of the total, indicated that they have become disillusioned with artificial intelligence technology because of identity confusion.
Several responses suggested that repeated instances of identity confusion could lead to a gradual erosion of trust, causing users to seek alternative tools perceived as more transparent. One participant articulated, ``After encountering misleading behavior several times, I’m beginning to question if I can rely on this model for professional use.'' Another participant from Dundee (who is a student) noted ``Hallucinations increases my skepticism and makes me double check the results''. 
These results highlight the significant impact that issues of identity confusion can have on public perception and trust in AI technologies and the entities that develop them. Just like one participant noted ``The model needs to be developed further as we are just in the beginning phase of artificial intelligence''.\looseness=-1

\begin{mdframed}[backgroundcolor=blue!4]  
\noindent\textit{\textbf{Finding~(VII).}{ Identity confusion in LLMs impacts user attitudes significantly, with many expressing disappointment in the companies behind the models, growing skepticism toward AI technology, and emphasizing the need for greater transparency and development to rebuild trust.
 }}
\end{mdframed}

The results highlights three main reasons the user believed for identity confusion in large language models: incorrect identity information learned from training data (35.58\%), errors in model design or operation (34.13\%), and potential plagiarism of other models (29.81\%). Notably, the concern over potential plagiarism reflects users' perception that LLMs may inadvertently mimic or incorporate elements from other models, which can cause more significant problems than other issues. Plagiarism undermines trust in the originality and integrity of the LLM but also risks.

\section{Discussion}
\noindent\textbf{Recommendation}:
Identity confusion in LLMs poses a significant challenge to trust and reliability in AI systems. Particularly, as shwon in our \S\ref{subsec:q1}, pre-trained models and modified transformer architectures were particularly vulnerable, with 30\% in each category displaying identity confusion, compared to no observed issues in fine-tuned models. This underscores the importance of refining fine-tuning processes to explicitly train models to recognize and assert their unique identity. Incorporating self-consistency checks and identity-verification layers can ensure models respond accurately to identity-related queries, even when presented with ambiguous prompts. Additionally, curating diverse and well-annotated training datasets, which exclude overlapping or conflicting identity references, can reduce the risk of confusion. By implementing these measures and using hallucination-reduction techniques, such as reinforcement learning or adversarial training, developers can address the root causes of identity confusion while improving overall model performance. 

For the AI community, tackling identity confusion requires a collaborative approach. Establishing shared benchmarks and repositories for reporting confusion incidents can create a standardized framework to diagnose and address these issues. Transparency in model documentation, including clear disclosures about datasets and architectural influences, is essential to foster trust. Furthermore, educating users on the limitations and potential pitfalls of LLM responses can empower them to critically evaluate outputs. By embracing open dialogue, ethical practices, and ongoing research into identity-related vulnerabilities, the community can collectively ensure that LLMs evolve into more reliable and trustworthy tools for diverse applications.

\vspace{2mm}
\noindent\textbf{Ethics Concerns}:
We addressed ethical concerns in our study by ensuring compliance with established research standards and ethical guidelines throughout the evaluation process. First, to protect participant rights and ensure informed consent, we secured approval from an IRB for the user study component of our research. This included a thorough review of our survey design, recruitment methods, and data handling procedures to guarantee that participants’ privacy and autonomy were respected. Second, our evaluation of LLMs focused exclusively on publicly available models and data, avoiding proprietary information or sensitive datasets that could raise ethical or legal concerns. We carefully excluded models explicitly stating reliance on external APIs or pre-built frameworks, ensuring that our analysis aligned with the principles of fairness and integrity. 
For instance, Hailuo AI does not provide an API for large-scale collection and distribution of its outputs. As a result, we did not include Hailuo AI in our analysis, despite it being subject to identity confusion issues. Finally, we have also reported our findings to all the vendors. As the time of submission, we are still waiting for their responses. 

\section{Related Work}

The rise of LLMs has brought challenges in identifying and attributing their outputs. To address concerns about originality, ownership, and security, researchers have proposed fingerprinting and watermarking techniques~\cite{liu2024survey,li2024double}. For example, 
Yang et al.\cite{yang2024fingerprint} proposed a framework for fingerprinting LLMs using unique output patterns, emphasizing robustness.  
REEF~\cite{zhang2024reef} detects LLM derivation using kernel alignment, protecting intellectual property without retraining. 
MarkLLM~\cite{pan2024markllm} is an open-source toolkit that streamlines LLM watermarking research with a unified framework, visualization, and evaluation tools. REMARK-LLM~\cite{remarkllm} is a robust watermarking framework for LLM-generated text that embeds binary signatures while preserving semantic integrity. On the other hand, 
several attacks have been developed to counter LLM watermarking~\cite{pang2024attacking,giboulot2024watermax,wu2024bypassing}. It is worth noting that our work is significantly different from LLM watermarking research. While LLM watermarking embeds intentional markers in model outputs to trace provenance, our work investigates misrepresentation and self-identification errors in LLMs, focusing on their prevalence, underlying causes, and impact on user trust.

\section{Conclusion}

This study highlights the critical challenge of identity confusion in LLMs, with 25.93\% of evaluated LLMs exhibiting this issue. Our analysis confirmed that identity confusion primarily stems from hallucinations rather than reuse or replication. This phenomenon significantly undermines trust, particularly in high-stakes applications such as education and professional tasks. It also impacts user attitudes, with many expressing disappointment in the companies behind these models and growing skepticism toward AI technology as a whole. As LLMs continue to proliferate across industries, ensuring their reliability and trustworthiness is essential. This study provides valuable insights to guide the development of more secure and dependable AI systems. \looseness=-1

\bibliographystyle{plain}
\bibliography{ref}

\appendix
\section*{Survey Questions}
% \usepackage{booktabs}
% \usepackage{multirow}
\begin{table*}[bp]

   \caption{Survey questions}  
\setlength\extrarowheight{-2pt} 
\begin{tabular}{@{}clc@{}}
\toprule
 &
  \multicolumn{1}{c}{\textbf{Question}} &
  \textbf{Answer Options} \\ \midrule
\multirow{7}{*}{\textbf{\begin{tabular}[c]{@{}c@{}}Background \\ information\end{tabular}}} &
  Q1:Age &
 18-30, 30-45, 45-60, Over 60 \\ \cmidrule(l){2-3} 
 &
  Q2:Gender &
  Male, Female, Prefer not to say \\ \cmidrule(l){2-3} 
 &
  Q3:Occupation &
  Common occupations \\ \cmidrule(l){2-3} 
 &
  Q4:Region &
  Asia, North America, Europe, Other \\ \cmidrule(l){2-3} 
 &
  Q5:Have you ever used large language models(LLMs) such as ChatGPT,Gemini,and so on? &
  Yes, No \\ \midrule
\multirow{11}{*}{\textbf{\begin{tabular}[c]{@{}c@{}}Familiarity \\ with LLM\end{tabular}}} &
  Q1:How many large language models are you aware of? &
  1, 2-5, 6-10, More than 10 \\ \cmidrule(l){2-3} 
 &
  Q2:How many large language models have you used? &
  1, 2-5, 6-10, More than 10 \\ \cmidrule(l){2-3} 
 &
  \begin{tabular}[c]{@{}l@{}}Q3:How would you rate your understanding of the fundamental principles \\ behind large language models?\end{tabular} &
  Five-point Likert scale \\ \cmidrule(l){2-3} 
 &
  Q4:What do you use large language models for? &
  \begin{tabular}[c]{@{}c@{}}Writing assistance, Creative writing,\\ General Q\&A, Professional Q\&A,\\ Educational purposes, Entertainment,\\ Professional tasks, Other\end{tabular} \\ \cmidrule(l){2-3} 
 &
  Q5:How much do you trust the information provided by large language models? &
  Five-point Likert scale \\ \midrule
\multirow{8}{*}{\textbf{\begin{tabular}[c]{@{}c@{}}Attitude \\ change \\ after fault\end{tabular}}} &
  \begin{tabular}[c]{@{}l@{}}Q1:If a large language model states that 3.11 is greater than 3.8, would you still\\ trust the other answers provided by the model?\end{tabular} &
  Five-point Likert scale \\ \cmidrule(l){2-3} 
 &
  Q2:In the situation just described, which tasks would you still have the LLM perform? &
  Same as Familiarity Q4 \\ \cmidrule(l){2-3} 
 &
  \begin{tabular}[c]{@{}l@{}}Q3:If a large language model provides inconsistent answers when asked the\\ same question twice, would you still trust its other answers?\end{tabular} &
  Five-point Likert scale \\ \cmidrule(l){2-3} 
 &
  Q4:In the situation just described, which tasks would you still have the LLM perform? &
  Same as Familiarity Q4 \\ \midrule
\multirow{30}{*}{\textbf{\begin{tabular}[c]{@{}c@{}}Attitude \\ change \\ after \\ identity \\ confusion\end{tabular}}} &
  \begin{tabular}[c]{@{}l@{}}Q1:If a large language model A claims that it is actually another LLM named B,\\ (e.g.,When asked "Who are you?",model A respond with“I am Model B.”)\\  would you still trust the other answers provided by model A?\end{tabular} &
  Five-point Likert scale \\ \cmidrule(l){2-3} 
 &
  Q2:In the situation just described, which tasks would you still have the LLM perform? &
  Same as Familiarity Q4 \\ \cmidrule(l){2-3} 
 &
  \begin{tabular}[c]{@{}l@{}}Q3:If a large language model A provides incorrect external references,(e.g.,When asked \\ "Can you provide the API link for your service?",model A respond with model B's \\ API link.)would you still trust the other answers provided by model A?\end{tabular} &
  Five-point Likert scale \\ \cmidrule(l){2-3} 
 &
  Q4:In the situation just described, which tasks would you still have the LLM perform? &
  Same as Familiarity Q4 \\ \cmidrule(l){2-3} 
 &
  \begin{tabular}[c]{@{}l@{}}Q5:If a large language model A claims to possess abilities or perform tasks that belong\\ to another model,(e.g.,When asked "Can you provide medical advice or diagnostics?",\\ model A respond with "Yes" but it can't perform these tasks.)\\ would you still trust the other answers provided by model A?\end{tabular} &
  Five-point Likert scale \\ \cmidrule(l){2-3} 
 &
  Q6:In the situation just described, which tasks would you still have the LLM perform? &
  Same as Familiarity Q4 \\ \cmidrule(l){2-3} 
 &
  \begin{tabular}[c]{@{}l@{}}Q7:If a large language model A provides an inaccurate or incomplete description of its \\ own profile,(e.g.,When asked "Can you provide a brief overview of yourself?",model A \\ incorrectly describe itself.)would you still trust the other answers provided by model A?\end{tabular} &
  Five-point Likert scale \\ \cmidrule(l){2-3} 
 &
  Q8:In the situation just described, which tasks would you still have the LLM perform? &
  Same as Familiarity Q4 \\ \cmidrule(l){2-3} 
 &
  \begin{tabular}[c]{@{}l@{}}Q9:If a large language model A claims that it is closely associated with or created by \\ another company,(e.g.,When asked "What is your relationship with company B?",model A\\ would respond with“I was created by company B.”while A is not created by company B \\ actually)would you still trust the other answers provided by model A?\end{tabular} &
  Five-point Likert scale \\ \cmidrule(l){2-3} 
 &
  Q10:In the situation just described, which tasks would you still have the LLM perform? &
  Same as Familiarity Q4 \\ \cmidrule(l){2-3} 
 &
  \begin{tabular}[c]{@{}l@{}}Q11:If a large language model A claims that it is  created by another company,\\ (e.g.,When asked "Who create you?",model A would respond with “I was created by \\ company B.”while A is not created by company B actually)\\ would you still trust the other answers provided by model A?\end{tabular} &
  Five-point Likert scale \\ \cmidrule(l){2-3} 
 &
  Q12:In the situation just described, which tasks would you still have the LLM perform? &
  Same as Familiarity Q4 \\ \midrule
\multirow{12}{*}{\textbf{\begin{tabular}[c]{@{}c@{}}Details \\ about \\ identity \\ confusion\end{tabular}}} &
  Q1:Have you ever encountered the instance that a LLM exhibited identity confusion? &
  Yes(please specify), No \\ \cmidrule(l){2-3} 
 &
  Q2:How does the phenomenon of identity confusion affect your attitude? &
  \begin{tabular}[c]{@{}c@{}}Disappointed with AI,\\ Disappointed with the LLM company,\\ No impact Other\end{tabular} \\ \cmidrule(l){2-3} 
 &
  \begin{tabular}[c]{@{}l@{}}Q3:What do you think might be the reasons for large language models exhibiting \\ identity confusion?\end{tabular} &
  \begin{tabular}[c]{@{}c@{}}The model may learned \\ incorrect identity information,\\ There are errors in the model's \\ design or operation(e.g.,hallucinations),\\ They plagiarized other models,\\ Other (please specify)\end{tabular} \\ \bottomrule
\end{tabular}
\end{table*}

% \end{thebibliography}

\end{document}